\documentclass[12pt]{article}

\usepackage{amsmath}
\usepackage{amssymb}
\usepackage{epsfig}
\usepackage{graphicx}
\usepackage{cite}
\usepackage{multirow}
\usepackage{longtable}
\usepackage{lscape}
\usepackage{bbm}
\usepackage{dcolumn}
\usepackage{bm}
\usepackage{booktabs}
\usepackage{rotating}
\usepackage{mathrsfs}

\allowdisplaybreaks[1] 

\newcommand{\capdef}{}
\newcommand{\mycaption}[2][\capdef]{\renewcommand{\capdef}{#2}%
        \caption[#1]{{\footnotesize #2}}}
\makeatletter
\renewcommand{\fnum@table}{\textbf{\tablename~\thetable}}
\renewcommand{\fnum@figure}{\textbf{\figurename~\thefigure}}
\makeatother

\newcounter{myenumi}

\renewcommand{\themyenumi}{\roman{myenumi}}
{\end{list}}

\newlength{\myem}
\newcounter{mysubequation}[equation]

\makeatletter
\renewcommand{\section}{\@startsection{section}{1}{0em}{-\baselineskip}%
{\baselineskip}{\normalfont\large\bfseries}}
\renewcommand{\subsection}%
{\@startsection{subsection}{2}{0em}{-0.7\baselineskip}%
{0.7\baselineskip}{\normalfont\bfseries}}
\makeatother

\newcommand{\bi}{\begin{itemize}}
\newcommand{\ei}{\end{itemize}}

\newcommand{\be}{\begin{equation}}
\newcommand{\ee}{\end{equation}}
\newcommand{\bea}{\begin{eqnarray}}
\newcommand{\eea}{\end{eqnarray}}

\newcommand{\msc}[1]{\mathscr{#1}}
\newcommand{\trm}[1]{\textrm{#1}}
\newcommand{\mcl}[1]{\mathcal{#1}}

\newcommand{\ie}{{\it i.e.}}

\newcommand{\eg}{{\it e.g.}}

\newcommand{\cf}{{\it cf.}}

\newcommand{\etc}{{\it etc.}}
\newcommand{\eq}{Eq.}

\newcommand{\fig}{Fig.}

\newcommand{\Ref}{Ref.}
\newcommand{\Refs}{Refs.}
\newcommand{\Sec}{Sec.}

\newcommand{\Tab}{Table}

\newcommand{\equ}[1]{\eq~(\ref{equ:#1})}
\newcommand{\figu}[1]{\fig~\ref{fig:#1}}

\begin{document}

\begin{titlepage}

\renewcommand{\thefootnote}{\alph{footnote}}

\vspace*{-3.cm}
\begin{flushright}
LPT-ORSAY-09-45 \\
FTUAM 09-17 \\
IFT-UAM/CSIC-09-36 
\end{flushright}


\renewcommand{\thefootnote}{\fnsymbol{footnote}}
\setcounter{footnote}{-1}

{\begin{center}
{\large\bf
Neutrino masses from higher than d=5 effective operators
} \end{center}}
\renewcommand{\thefootnote}{\alph{footnote}}

\vspace*{.8cm}
\vspace*{.3cm}
{\begin{center} {\large{\sc
                F.~Bonnet\footnote[1]{\makebox[1.cm]{Email:}
                florian.bonnet@th.u-psud.fr, florian.bonnet@pd.infn.it},
 		D.~Hernandez\footnote[2]{\makebox[1.cm]{Email:}
                hernandez@delta.ft.uam.es}, 
 		T.~Ota\footnote[3]{\makebox[1.cm]{Email:}
                toshihiko.ota@physik.uni-wuerzburg.de}, and
                W.~Winter\footnote[4]{\makebox[1.cm]{Email:}
                winter@physik.uni-wuerzburg.de}
                }}
\end{center}}
\vspace*{0cm}
{\it
\begin{center}

\footnotemark[1]Laboratoire de Physique Th\'eorique UMR 8627,\\
 Universit\'e Paris-Sud 11, Bat. 210, 91405 Orsay Cedex, France

\footnotemark[2]
	Departamento de F{\'i}sica Te{\'o}rica and Instituto de F{\'i}sica Te{\'o}rica UAM/CSIC, \\
        Universidad Aut{\'o}noma de Madrid, 28049 Cantoblanco, Madrid, Spain 

\footnotemark[3]${}^,$\footnotemark[4]
       Institut f{\"u}r Theoretische Physik und Astrophysik, \\ Universit{\"a}t W{\"u}rzburg, 
       97074 W{\"u}rzburg, Germany

\end{center}}

\vspace*{1.5cm}

{\Large \bf
\begin{center} Abstract \end{center}  }

We discuss the generation of small neutrino masses from effective operators higher than dimension five, which open new possibilities for low scale see-saw mechanisms. In order to forbid the radiative generation
of neutrino mass by lower dimensional operators, extra fields are required, which are charged under a new symmetry.
We discuss this mechanism in the framework of a two Higgs doublet model. We demonstrate that the tree level generation of neutrino mass from higher dimensional operators often leads to inverse see-saw scenarios in which small lepton number violating terms are naturally suppressed by the new physics scale. Furthermore, we systematically discuss tree level generalizations of the standard see-saw scenarios from higher dimensional operators. Finally, we point out that higher dimensional operators can also be generated at the loop level. In this case, we obtain the TeV scale as new physics scale even with order one couplings.

\vspace*{.5cm}

\end{titlepage}

\newpage

\renewcommand{\thefootnote}{\arabic{footnote}}
\setcounter{footnote}{0}


\section{Introduction}


The evidence for neutrino oscillations  can only be understood in terms of massive neutrinos (see \Ref~\cite{GonzalezGarcia:2007ib} for a recent review). In the Standard Model (SM) of elementary particle physics, however, neutrinos are massless. The simplest extensions of the SM model
to accommodate neutrino masses are additions of Dirac or Majorana neutrino mass terms. In order to have
Dirac mass terms, right-handed neutrinos are required, which are not present in the SM field content.
For Majorana neutrino masses, lepton number, which is an accidental symmetry of the SM, needs to be violated. 
Therefore, massive neutrinos require physics beyond the SM.

The current upper bound on neutrino masses of the order $1 \,
\mathrm{eV}$ suggests that the neutrinos are much lighter than their
charged SU(2) counterparts. As far as the different generations are
concerned, the hierarchy among neutrinos seems to be different from
those present in the other families of charged fermions, even in the
hierarchical case. In addition, compared to the quarks, leptons exhibit
strong generation mixing. 
Therefore, theories of neutrino mass are expected to account for their smallness \emph{as well as} for the flavor structure in the lepton sector. Though we intend to focus on the former issue in this work, the models we consider leave room to be extended by a successful theory of flavor. 

Since neutrino mass requires physics beyond the SM, it is convenient to parameterize the impact of the heavy fields, present in the high-energy theory,  by the addition  of a tower  of effective  operators $\mathcal{O}^{d}$ of dimension $d>4$ to the Lagrangian. These operators are made out of the SM fields, are invariant under the SM gauge group~\cite{Weinberg:1979sa,Wilczek:1979hc} (see also \Ref~\cite{Buchmuller:1985jz}) 
 and are {\it non-renormalizable}. They parameterize the effects of the high energy degrees of freedom on the low energy theory order by order.
The operator coefficients are weighted by inverse powers of the scale of new physics $\Lambda_{\mathrm{NP}}$: 
 \be\label{equ:leff}
\mathscr{L} = \mathscr{L}_{\rm SM} + \mathscr{L}^{d=5}_{\text{eff}} 
+ \mathscr{L}^{d=6}_{\text{eff}} + \cdots
\, , \quad \textrm{with} \quad \mathscr{L}^{d}_{\text{eff}} \propto \frac{1}{\Lambda_{\mathrm{NP}}^{d-4}} \, \mathcal{O}^{d}\,.
\ee
Some of these effective operators result in corrections to the
 low-energy SM parameters and in exotic couplings. As an example,
 consider the well known case of lepton number conserving operators built only with lepton fields and the Higgs. These lead, at $d=6$, $d=8$, \etc, to charged lepton flavor violation and non-standard neutrino interactions\cite{Berezhiani:2001rs,Davidson:2003ha,Antusch:2008tz,Gavela:2008ra,Biggio:2009kv,Biggio:2009nt}.

It is also known that there is only one possible operator at the lowest order in the expansion, $\msc{L}^{d=5}_{\trm{eff}}$,  namely, the famous Weinberg operator~\cite{Weinberg:1979sa},
\be
\mcl{O}_W = (\overline{L^{c}} {\rm i} \tau^{2} H)\, (H {\rm i} \tau^2 L)
\ee
which leads, after Electroweak Symmetry Breaking (EWSB), to Majorana masses  for the neutrinos (Here $L$ and $H$ stand for the Standard Model lepton doublets and Higgs field, respectively).  At tree level, $\mcl{O}_W$ can only be mediated by a singlet fermion, a triplet scalar, or a triplet fermion, leading to the famous type~I~\cite{Minkowski:1977sc,Yanagida:1979as,GellMann:1980vs,Mohapatra:1979ia}, type~II~\cite{Magg:1980ut,Schechter:1980gr,Wetterich:1981bx,Lazarides:1980nt,Mohapatra:1980yp,Cheng:1980qt}, and type~III\cite{Foot:1988aq} see-saw formulae, respectively (see also \Ref~\cite{Abada:2007ux}). Compared to the electroweak scale, the mass of the neutrinos in all three cases appears suppressed by a factor 
$v/\Lambda_{\mathrm{NP}}$, where $v/\sqrt{2}$ is the Vacuum Expectation Value (VEV) of the Higgs. Substituting typical values, one obtains that the original see-saw mechanisms point towards the GUT scale.

More recently, however, scenarios in which $\Lambda_{\mathrm{NP}} \sim
\mathrm{TeV}$ have drawn some attention since they are potentially
testable at the LHC and future neutrino facilities
(see, \eg,
\Refs~\cite{Keung:1983uu,Langacker:1988up,Langacker:1988ur,Tommasini:1995ii,Flanz:1999ah,Czakon:2001em,Broncano:2002rw,Broncano:2003fq,Pilaftsis:2005rv,Bajc:2006ia,Kersten:2007vk,Bray:2007ru,delAguila:2007em,Abada:2007ux,Garayoa:2007fw,Holeczek:2007kk,FernandezMartinez:2007ms,Bajc:2007zf,Goswami:2008mi,Bartl:2009an,Altarelli:2008yr,Antusch:2009pm,Malinsky:2009gw,Malinsky:2009df,Arhrib:2009mz}
for their phenomenology). In  these cases, additional suppression
mechanisms for the neutrino masses are required, and several
possibilities open up:
\begin{enumerate}
\item
     The neutrino mass is generated radiatively. The additional suppression
     is guaranteed by the loop
     integrals~\cite{Zee:1980ai,Wolfenstein:1980sy,Zee:1985rj,Babu:1988ki,Ma:1998dn,FileviezPerez:2009ud,Babu:2002uu,Krauss:2002px,Cheung:2004xm,Ma:2006km,Ma:2007yx,Aoki:2008av,Aoki:2009vf}.

\item 
      The neutrino mass is generated at tree level, where additional
      suppression enters through a small lepton number violating
      contribution (\eg, in inverse see-saw scenarios, R parity-violating
      SUSY models and so on~\cite{Schechter:1981bd,Nandi:1985uh,Mohapatra:1986bd,Branco:1988ex,GonzalezGarcia:1988rw,Xing:2009hx,Ma:2000cc,Tully:2000kk,Loinaz:2003gc,Hirsch:2004he,Pilaftsis:2005rv,Gouvea:2007xp,Kersten:2007vk,Grimus:2009mm}).

\item
     The neutrino mass is forbidden at $d=5$, but appears from 
     effective operators of higher dimension~\cite{Babu:1999me,Chen:2006hn,Gogoladze:2008wz,Giudice:2008uua,Babu:2009aq,Gu:2009hu}.
\end{enumerate}
All these cases lead to neutrino masses suppressed by a scale of new physics  much smaller 
than the GUT scale and potentially as small as the TeV scale. In this study, we focus on the third possibility, that is, we will assume that neutrino masses come from effective operators of dimension higher than $d=5$ as a starting point.
However, as we will demonstrate in \Sec~\ref{sec:invdecomp}, it is possible to find scenarios which
naturally contain the second option with the lepton number violating parameter suppressed by $1/\Lambda_{\mathrm{NP}}$. In addition, we will show in \Sec~\ref{sec:higher} how the three ideas can be combined, to obtain the TeV scale even with order one couplings.

Disregarding flavor, spinor, and gauge indices, 
the lepton number violating $d=5$, $d=7$, \etc, operators in \equ{leff} that contribute to Majorana neutrino masses are of the form
\begin{eqnarray}
\mathcal{O}^5 & = & \mcl{O}_W =  LLHH \label{equ:d5}  \\
\mathcal{O}^7 & = & (LLHH)(H^\dagger H) \label{equ:d7}  \\
\mathcal{O}^9 & = & (LLHH)(H^\dagger H)(H^\dagger H) \label{equ:d9}  \\
& \vdots & \nonumber
\end{eqnarray}
If we assume that the neutrino Yukawa coupling is naturally $\mathcal{O}(1)$, then the masses of the neutrinos are roughly given by
\begin{equation}
m_\nu \sim v \left( \frac{v}{\Lambda_{\mathrm{NP}}} \right)^{d-4} \, .
\label{equ:numass}
\end{equation}
For a typical neutrino mass of the order of electron volt, this relationship gives the energy scale of new physics as a function of the dimension $d$ of the operator responsible for neutrino masses. 
If we want to lower the scale of new physics down to that of present or near future experiments, $\Lambda_{\mathrm{NP}} \sim 1-10$ TeV, then $d \geq 9$ suffices in case no additional suppression mechanism is provided. On the other hand, if Yukawas of the order $m_e/v \simeq 10^{-6}$ are considered natural, then $d \geq 7$ is enough.

It is worth to take a closer look at the complications involved. For that, let $D$ be the dimension of the operator that gives the dominant contribution to neutrino masses. In order to claim $D>5$, we need all relevant operators of dimension $d<D$  to be forbidden.
Indeed, taking for instance the operator in 
\equ{d7}, it is clear that the $(H^\dagger H)$ component can be closed in a loop. This leads to the $d=5$ Weinberg operator with the additional suppression factor $1/(16 \pi^2)$ -- unless the loop contributions cancel: 
 \begin{equation}
 \frac{1}{\Lambda_{\mathrm{NP}}^3} (LLHH)(H^\dagger H) \rightarrow \frac{1}{16 \pi^2} \, \frac{1}{\Lambda_{\mathrm{NP}}} \, (LLHH) \, .
\end{equation}
 The latter will be the leading contribution to neutrino masses if
 $1/(16 \pi^2) \gtrsim (v/\Lambda_{\mathrm{NP}})^2$, that is, if
 $\Lambda_{\mathrm{NP}} \gtrsim 3 \, \mathrm{TeV}$. 
Note that in both cases $\Lambda_{\rm NP} \lesssim 3$ TeV and 
$\Lambda_{\rm NP} \gtrsim 3$ TeV, the new physics might have
implications at the LHC.
For a robust model valid in the entire LHC-testable range,
one should therefore have a fundamental reason
--- symmetry --- to justify the leading contribution to neutrino mass.

We call a dimension $d \geq 7$ operator \emph{genuine} if it is impossible to generate  some other neutrino mass operator of lower dimension by closing loops. In this work, we seek for a genuine operator, which means that we need a symmetry that forbids the appearance of neutrino masses at dimension $d < D$. One can easily see from Eqs.~(\ref{equ:d5}) to~(\ref{equ:d9}) that the symmetry cannot be implemented with SM fields only. This is because the combination $(H^\dagger H)$ is a singlet under any symmetry and therefore, if one operator is allowed, then the whole tower must be so. On the contrary, one can slightly enlarge  the Higgs sector and charge the fields under a  new  U(1) or discrete symmetry (a so-called ``matter parity''\cite{Ibanez:1991pr}) that allows a dimension $D$ operator while forbidding all others with lower dimension.

In this context, the simplest possibilities to enhance the field content of the SM are the addition of a
Higgs singlet~\cite{Chen:2006hn,Gogoladze:2008wz}
\begin{align}
 \mathscr{L}^{d=n+5}_{\text{eff}} =  \frac{1}{\Lambda_{\mathrm{NP}}^{d-4}} (LLHH) (S)^{n} \, , \quad n=1,2,3, \hdots 
\label{equ:s}
\end{align}
or the addition of a Higgs doublet, leading to the Two Higgs Doublet Model (THDM)~\cite{Gunion:1989we,Babu:1999me,Giudice:2008uua}
\begin{align} 
\mathscr{L}^{d=2n+5}_{\text{eff}} =  \frac{1}{\Lambda_{\mathrm{NP}}^{d-4}}
	(LLH_{u} H_{u}) (H_{d} H_{u})^{n} \, , \quad n=1,2,3, \hdots \, .
\label{equ:hh}
\end{align}
More complicated options include, for instance, the next-to-minimal SUSY standard model (NMSSM) using two Higgs doublets and a scalar, see \Ref~\cite{Gogoladze:2008wz}. In this study, we only consider \equ{hh} within the THDM. However, as we shall discuss elsewhere~\cite{Gavela:prep}, our mechanism can be applied to SUSY models as well.

We discuss in \Sec~\ref{sec:eff} the conditions to obtain neutrino masses from genuine effective operators of dimension $d \geq 7$. Then we show in \Sec~\ref{sec:invdecomp} several tree level decompositions of the only $d=7$ operator allowed in both SUSY and the THDM, which describe the smallness of the lepton number violating terms naturally. Furthermore, we discuss generic extensions of the standard see-saw scenarios in \Sec~\ref{sec:sys}, and we illustrate additional suppression mechanisms, such as from even higher dimensional operators or loop suppression factors, in  \Sec~\ref{sec:higher}.

\section{Neutrino mass from higher dimensional operators}
\label{sec:eff}

In order to have a \emph{genuine} dimension $D$ operator to be the leading contribution to neutrino mass, we forbid all $d<D$ operators by means of a new U(1) or $\mathbb{Z}_n$ symmetry. We assign matter charges $q$ (see, \eg, \Refs~\cite{Ibanez:1991pr,Shirai:2009fq}) to the new fields $H_u$ ($q_{H_u}$), $H_d$ ($q_{H_d}$), and the SM fields, \ie, the lepton doublets $L$ ($q_L$), right-handed charged leptons $E$ ($q_E$), quark doublets $Q$ ($q_Q$), right-handed up-type quarks $U$ ($q_U$), and right-handed down-type quarks $D$ ($q_D$). 

For the following discussion, we show the charge assignments assuming a discrete $\mathbb{Z}_n$ symmetry. Note, however, that the effective operators can be controlled as well by a new U(1) symmetry. If that is the case, additional (unwanted) Goldstone bosons may appear after the spontaneous breaking of the electroweak and U(1) symmetry. As we will discuss later, this can be avoided by breaking the U(1) explicitly, either by an enhanced scalar sector, or by a soft breaking term. Since the actual implementation of this U(1) breaking depends on the model, we will not touch it in this section, and focus on the discrete symmetries for the moment.

\begin{table}[t!]
\begin{tabular}{cccccl}
\hline \hline
& SUSY
& Op.\#
&Effective interaction
& Cond.\# & Charge of effective int.\\
\hline
dim.5
& $\checkmark$
& 1
& $LLH_{u} H_{u}$
& 1 
& $2q_{L} + 2q_{H_{u}}$
\\
\cline{2-6}
&
& 2
&$LLH_{d}^{*} H_{u}$ 
& 2
& $2q_{L} - q_{H_{d}} + q_{H_{u}}$
\\
&
& 3
& $LLH_{d}^{*} H_{d}^{*}$ 
& 3
& $2q_{L} - 2 q_{H_{d}}$
\\
\hline
dim.7
&  $\checkmark$
& 4
& $LLH_{u} H_{u} H_{d} H_{u}$
& 4
& $2 q_{L} + q_{H_{d}} + 3 q_{H_{u}}$
\\
\cline{2-6}
&
& 5
& $LLH_{u} H_{u} H_{d}^{*} H_{d}$ 
& 1
& $2q_{L} + 2q_{H_{u}}$
\\
& 
& 6
& $LLH_{u} H_{u} H_{u}^{*} H_{u}$ 
& 1
& $2q_{L} + 2q_{H_{u}} $
\\
&
& 7
& $LLH_{d}^{*} H_{u} H_{d}^{*} H_{d} $ 
& 2
& $2q_{L} - q_{H_{d}} + q_{H_{u}}$
\\
&
& 8
& $LLH_{d}^{*} H_{u} H_{u}^{*} H_{u} $ 
& 2
& $2q_{L} - q_{H_{d}} + q_{H_{u}}$
\\
&
& 9
& $LLH_{d}^{*} H_{d}^{*} H_{d}^{*} H_{d} $ 
& 3
& $2q_{L} - 2q_{H_{d}} $
\\
&
& 10
& $LLH_{d}^{*} H_{d}^{*} H_{u}^{*} H_{u} $ 
& 3
& $2q_{L} - 2q_{H_{d}} $
\\
&
& 11
& $LLH_{d}^{*} H_{d}^{*} H_{u}^{*} H_{d}^{*} $ 
& 5
& $2q_{L} - 3 q_{H_{d}} - q_{H_{u}}$
\\
\hline
dim.9
&  $\checkmark$
& 12
&  $LLH_{u} H_{u} H_{d} H_{u} H_{d} H_{u}$
& 6
& $2q_{L} + 2 q_{H_{d}} + 4 q_{H_{u}}$
\\
\cline{2-6}
&
& 13
& $LLH_{u} H_{u} H_{d} H_{u} H_{d}^{*} H_{d}$ 
& 4
& $2q_{L} + q_{H_{d}} + 3q_{H_{u}}$
\\
&
& 14
& $LLH_{u} H_{u} H_{d} H_{u} H_{u}^{*} H_{u}$ 
& 4
& $2q_{L} + q_{H_{d}} + 3q_{H_{u}}$
\\
&
& 15
& $LLH_{u} H_{u} H_{d}^{*} H_{d} H_{d}^{*} H_{d}$ 
& 1
& $2q_{L} + 2q_{H_{u}}$
\\
&
& 16
& $LLH_{u} H_{u} H_{d}^{*} H_{d} H_{u}^{*} H_{u}$ 
& 1
& $2q_{L} + 2q_{H_{u}}$
\\
&
& 17
& $LLH_{u} H_{u} H_{u}^{*} H_{u} H_{u}^{*} H_{u}$ 
& 1
& $2q_{L} + 2q_{H_{u}}$
\\
&
& 18
& $LLH_{d}^{*} H_{u} H_{d}^{*} H_{d} H_{d}^{*} H_{d}$ 
& 2
& $2q_{L} - q_{H_{d}} + q_{H_{u}}$
\\
&
& 19
& $LLH_{d}^{*} H_{u} H_{d}^{*} H_{d} H_{u}^{*} H_{u}$ 
& 2
& $2q_{L} - q_{H_{d}} + q_{H_{u}}$
\\
&
& 20
& $LLH_{d}^{*} H_{u} H_{u}^{*} H_{u} H_{u}^{*} H_{u}$ 
& 2
& $2q_{L} - q_{H_{d}} + q_{H_{u}}$
\\
&
& 21
& $LL H_{d}^{*} H_{d}^{*} H_{d}^{*} H_{d} H_{d}^{*} H_{d}$ 
& 3
& $2q_{L} - 2q_{H_{d}} $
\\
&
& 22
& $LL H_{d}^{*} H_{d}^{*} H_{d}^{*} H_{d} H_{u}^{*} H_{u}$ 
& 3
& $2q_{L} - 2q_{H_{d}} $
\\
&
& 23
& $LL H_{d}^{*} H_{d}^{*} H_{u}^{*} H_{u} H_{u}^{*} H_{u}$ 
& 3
& $2q_{L} - 2q_{H_{d}} $
\\
&
& 24
& $LLH_{d}^{*} H_{d}^{*} H_{d}^{*} H_{u}^{*} H_{d}^{*} H_{d}$ 
& 5
& $2q_{L} - 3q_{H_{d}} - q_{H_{u}}$ 
\\
&
& 25
& $LLH_{d}^{*} H_{d}^{*} H_{d}^{*} H_{u}^{*} H_{u}^{*} H_{u}$ 
& 5
& $2q_{L} - 3q_{H_{d}} - q_{H_{u}}$ 
\\
&
& 26
&$LLH_{d}^{*} H_{d}^{*} H_{u}^{*} H_{d}^{*} H_{u}^{*} H_{d}^{*}$ 
& 7
& $2q_{L} - 4q_{H_{d}} - 2 q_{H_{u}}$ \\
\hline
dim.11 & & & $\hdots$  \\
\hline \hline
\end{tabular}
\mycaption{\label{tab:operators}
Effective operators generating
 neutrino mass in the THDM.
 In SUSY models, only the operators with the column ``SUSY''
checked are allowed because of the holomorphy of super-potential.
In the last two columns, we show the charge of the effective interaction
with respect to our discrete symmetry, and we number the independent
conditions.
}
\end{table}

We list the possible $d=5$, $d=7$, and $d=9$ effective operators that generate neutrino mass together with the charge of the effective interaction in
\Tab~\ref{tab:operators}. Obviously, not all of the charges are independent, which we illustrate by giving each independent condition a number (second-last column). Genuine operators are precisely the ones whose charge is independent from all those of lower dimension. For instance, at order $d=7$, the only possible genuine operators  are \#4 and \#11. At $d=9$, there are again only two possibilities, operators \#12 and \#26.
In the following, we will use operator~\#4 as an example, since it is the simplest realization of our mechanism which is allowed in both the THDM and SUSY. Note that in SUSY models, only the operators with the column ``SUSY''
checked are allowed because of the holomorphy of super-potential.

In order to have operator~\#4 as leading contribution, we need to allow this operator by the condition on the $\mathbb{Z}_n$ charges
\begin{align}
(2 q_{L} + q_{H_{d}} + 3 q_{H_{u}}) \mod n = 0&
\label{eq:cond-dim7-4}
\end{align}
and suppress all lower dimensional operators and all other $d=7$ operators by charging them as (\cf, \Tab~\ref{tab:operators}):
\begin{align}
(2q_{L} + 2q_{H_{u}}) \mod n \neq 0& 
\, ,
\label{eq:cond-dim7-1}
\\
(2q_{L} - q_{H_{d}} + q_{H_{u}}) \mod n \neq 0 &
\, ,
\label{eq:cond-dim7-2}
\\
(2q_{L} - 2 q_{H_{d}}) \mod n \neq 0 &
\, ,\label{eq:cond-dim7-3}
\\
(2q_{L} - 3 q_{H_{d}} - q_{H_{u}}) \mod n  \neq 0 &
\, .
\label{eq:cond-dim7-5}
\end{align}
In addition, we have to allow the ordinary Yukawa interactions which requires\footnote{%
Note that $E = (e_{R})^{c}$, $U = (u_{R})^{c}$, and $D = (d_{R})^{c}$.
We assume Yukawa interactions of the THDM type II (and MSSM)
in which Higgs-mediated flavour changing neutral current processes 
are suppressed~\cite{Gunion:1989we}.
}
\begin{align}
(q_{E} + q_{L} + q_{H_{d}}) \mod n = 0  \, ,
\label{eq:cond-Yukawa-e}
\\
(q_{D} + q_{Q} + q_{H_{d}}) \mod n = 0\, ,
\label{eq:cond-Yukawa-d}
\\
(q_{U} + q_{Q} + q_{H_{u}}) \mod n = 0  \, .
\label{eq:cond-Yukawa-u}
\end{align}  
Without loss of generality, we fix the charge of the quark doublet to be $q_Q=0$.

We have tested all possibilities for charge assignments and discrete symmetries systematically
in order to identify the simplest possibility in terms of group order (we do not consider group products). It has turned out that a $\mathbb{Z}_5$ symmetry is the simplest one, with, for instance, the following charge assignments 
\begin{equation}
q_{H_u} = 0 \, , \, \,  q_{H_d}=3 \, , \, \, q_L=1 \, , \, \,  q_E=1 \, , \, \,  q_Q = 0 \, , \, \, q_U= 0 \, , \, \,  q_D=2 \, \, . 
\label{equ:charges}
\end{equation}
For operator \#11, we also obtain $\mathbb{Z}_5$ as option with the lowest group order. For the $d=9$ operators \#12 and \#26, we need at least a $\mathbb{Z}_7$. If SUSY is implemented, both operators~\#4 and \#12  can be realized within a $\mathbb{Z}_3$. Note that the charge assignments are not unique.\footnote{There are $3 \times 2=6$ possibilities for $\mathbb{Z}_3$, $5 \times 4=20$ possibilities for $\mathbb{Z}_5$, $42 = 7 \times 6$ possibilities for $\mathbb{Z}_7$, \etc, because the first assignment is always arbitrary, the second is also arbitrary but cannot be hypercharge (one possibility subtracted), and the rest is determined by these two.}

From the discussion above, it should be clear that these operators can be generated at tree level, which we consider in the following two sections. The discrete symmetry (matter parity), which we have introduced, must be broken by the Higgses taking their VEVs, because the effective Majorana mass terms obviously violate the $\mathbb{Z}_5$. Note, however, that this symmetry is not the same as lepton number. This can easily seen by the effective operator $(\#1)^5$ made from operator~\#1 in \Tab~\ref{tab:operators}. This operator is obviously invariant under the $\mathbb{Z}_5$, but it violates lepton number. 


\section{Inverse see-saw mechanisms with naturally suppressed lepton number violation}
\label{sec:invdecomp}

\begin{figure}[t!]
\begin{center}
\unitlength=1cm
\begin{picture}(6,6)
\put(0,0){\includegraphics[width=6cm]{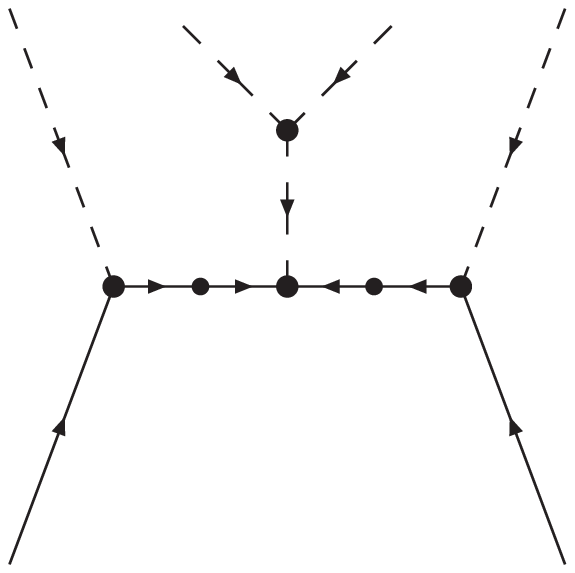}}
\put(-0.1,0){$L$}
\put(5.8,0){$L$}
\put(0,5.8){$H_{u}$}
\put(5.7,5.8){$H_{u}$}
\put(1.7,5.6){$H_{d}$}
\put(4,5.6){$H_{u}$}
\put(2.6,3.6){$\phi$}
\put(1.4,2.4){$N_{R}$}
\put(2.4,2.4){$N_{L}'$}
\put(4.1,2.4){$N_{R}$}
\put(3.3,2.4){$N_{L}'$}
\put(3,-0.8){(a)}
\end{picture}
\hspace{1.5cm}
\begin{picture}(6,6)
\put(0,0){\includegraphics[width=6cm]{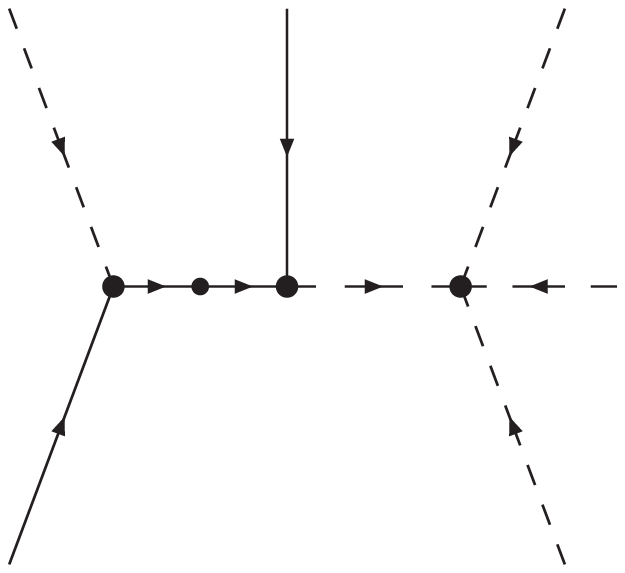}}
\put(-0.1,0){$L$}
\put(5.4,0){$H_{u}$}
\put(0,5.3){$H_{u}$}
\put(2.6,5.3){$L$}
\put(5.3,5.3){$H_{u}$}
\put(5.8,2.5){$H_{d}$}
\put(1.4,2.2){$N_{R}$}
\put(2.2,2.2){$N_{L}'$}
\put(3.5,2.2){$\Phi$}
\put(3,-0.8){(b)}
\end{picture}
\vspace*{1cm}
\end{center}
\mycaption{\label{fig:decomp} Tree level decompositions of the dimension seven 
 operator $LLH_{u} H_{u} H_{d} H_{u}$ (\#4 in \Tab~\ref{tab:operators})
for neutrino masses
leading to inverse see-saw scenarios with entries in
the (3,3) element (a) and (1,3) element (b). Here $N_{L}'$ and
$N_{R}$ refer to SU(2) singlet fermions, $\phi$ to a singlet scalar,
and $\Phi$ to a doublet scalar.}
\end{figure}

So far we have only discussed the effective operators and the necessary conditions to have a genuine $d>5$ operators as leading contribution for neutrino mass. We show in this section several examples to illustrate the completions of the theory at high energies. 

We consider see-saw-like models of the fermionic type. It is easy to convince oneself that the simplest cases, such as the type~I see-saw, can not produce a genuine $D \geq 7$ operator.\footnote{The type~I see-saw implies the introduction of the right-handed heavy Majorana mass term and the Yukawa interaction with the lepton doublet. If the Yukawa interaction is present in the theory, the right-handed Majorana mass term has to be obviously forbidden, because otherwise the usual $d=5$ operator is generated. Without the Majorana mass, however, no suppression is obtained. Therefore, at least one more fermionic field is required, and the fermionic fields need to form a mass term.}
Hence, we focus on tree level decompositions of $d=7$ operators which
require the addition of two extra fermion singlet fields ${N_R}_{a}$ and
${N'_L}_{a}$. 
This leads to an inverse see-saw-like
structure~\cite{GonzalezGarcia:1988rw,Mohapatra:1986bd,Nandi:1985uh} of
the neutral fermion mass matrix of the form (in the basis 
$\begin{pmatrix}
\nu_L^{c}
&
N_R
&
{N'_L}^{c}
\end{pmatrix}$):
\begin{equation}
 M_\nu = 
\begin{pmatrix}
 0 
 & (Y_{\nu}^{\sf T}) v
 & \epsilon (Y'^{\sf T}_{\nu})
 \\
 (Y_{\nu}) v
 & \mu'
 & \Lambda_{\mathrm{NP}}
 \\
 \epsilon (Y'_{\nu})
 & \Lambda_{\mathrm{NP}}^{\sf T}
 & \mu
\end{pmatrix} \, .
\label{equ:iss}
\end{equation}
Here $\epsilon$, $\mu$, and $\mu'$ are typically assumed to be small parameters, because they are responsible of the lepton number violation (for the $\epsilon$-term, see also \Refs~\cite{Abada:2007ux,Gavela:2009cd}).  

With only $N_R$ and $N_L'$ added to the SM, the interactions leading to the mass matrix in \equ{iss} can only be obtained via non-renormalizable operators. Indeed, assuming the charge assignments in \equ{charges}, the only renormalizable term that can be written is
\begin{equation}
\overline{N_R}Y_{\nu} H_u {\rm i}\tau^2 L+\textrm{H.c.}\,
\end{equation}
with $q_{N_{R}} = q_{N_{L}'} = 1$ in order to conserve the $\mathbb{Z}_5$ symmetry. Now constructing, with the SM fields plus $N_R$ and $N'_L$,  the possible effective operators that respect the $\mathbb{Z}_5$ symmetry, one obtains the $d=5$ operators
\begin{equation}
\frac{\lambda_1}{\Lambda_{\trm{NP}}}(H_d {\rm i}\tau^2 H_u)
 \overline{N_R}N_R^{c}
 +
 \frac{\lambda'_1}{\Lambda_{\trm{NP}}}(H_d {\rm i}\tau^2 H_u)
 \overline{N_{L}'^c}N'_L +\textrm{H.c.},
\label{equ:d=5iss}
\end{equation}
and the $d=6$ operator
\begin{equation}
\frac{\lambda_2}{\Lambda^2_{\trm{NP}}}
(H_d {\rm i} \tau^2 H_u)\overline{N_L'^{c}} Y_{\nu}' H_u {\rm i} \tau^2 L+\textrm{H.c.},
\label{equ:d=6iss}
\end{equation}
where $\Lambda_{\mathrm{NP}}$ is the higher energy scale. Matching these with Eq.~(\ref{equ:iss}) leads to 
\begin{equation}
\mu=\frac{\lambda_1}{\Lambda_{\trm{NP}}}\langle H_{d}^{0} H_{u}^{0} \rangle,\quad \mu'=\frac{\lambda'_1}{\Lambda_{\trm{NP}}}\langle H_{d}^{0} H_{u}^{0}\rangle,\quad \epsilon=\frac{\lambda_2}{\Lambda^2_{\trm{NP}}}\langle H_{d}^{0}\rangle {\langle H_{u}^{0}\rangle}^2 \, .
\end{equation}
 In order to generate those coefficients through renormalizable interactions, extra scalar fields need to be added, which masses will define $\Lambda_{\trm{NP}}$. From Eqs.~(\ref{equ:d=5iss}) and (\ref{equ:d=6iss}), it is clear that, at tree-level, the same field that generates the operators in Eq.~(\ref{equ:d=5iss}) cannot generate the operator in Eq.~(\ref{equ:d=6iss}). In other words, depending on the decomposition, the $\mu$-term or the $\epsilon$-term will be generated, but not both simultaneously.

In the following, we will show fundamental theories which predict a small lepton number violating (LNV) $\mu$- or $\epsilon$-term suppressed by the new physics scale. The diagrams generating neutrino mass are shown in \figu{decomp}.

\subsection{Decomposition (a): The $\boldsymbol{\mu}$-term}

For the decomposition (a) in \figu{decomp}, we introduce two chiral
fermions, singlets of the SM: 
$N_{R}$ (right-handed) and $N_{L}'$ (left-handed), 
and a SM singlet scalar $\phi$. 
The relevant interactions are then given by
\begin{align}
\mathscr{L} 
=&
\mathscr{L}_{\rm SM}
+
\left[
(Y_{\nu})_{a\alpha}\,
 (\overline{N_{R}})_{a}
 H_{u} 
 {\rm i} \tau^{2} 
 L_{\alpha}
 +
\kappa_{ab} 
(\overline{N_{L}'^{c}})_{a} (N_{L}')_{b} \phi 
+
\mu
 \phi^*
 H_{d} {\rm i} \tau^{2} H_{u}
 +
 (\overline{N_{R}})_{a}\,M_{ab}\,(N_{L}')_{b}
 +
 {\rm H.c.}
 \right]
\nonumber \\
&
+
 M_{\phi}^{2}
 \phi^{*} \phi
 +
 \cdots.
\label{equ:L-decon-3}
\end{align}
The mass matrix for the neutral fermion fields can be summarized  as
\begin{align}
\mathscr{L}=
\frac{1}{2}
\begin{pmatrix}
\overline{\nu_{L}^{c}}_{\alpha}
&
\overline{N_{R}}_a
&
(\overline{N_{L}'^{c}})_c
\end{pmatrix}
\begin{pmatrix}
 0 
 & (Y_{\nu}^{\sf T})_{\alpha b}\langle H_{u}^{0} \rangle 
 & 0 
 \\
 (Y_{\nu})_{a\beta} \langle H_{u}^{0} \rangle 
 & 0 
 & M_{ad} 
 \\
 0 
 & (M^{\sf T})_{cb}
 & (\Lambda^{-1})_{cd} \langle H_{d}^{0} H_{u}^{0} \rangle
\end{pmatrix}
\begin{pmatrix}
\nu_{L \beta} \\ (N_{R}^{c})_{b} \\ N_{L d}'
\end{pmatrix}
+{\rm H.c.},
\label{equ:massmatrix-inverseseesaw}
\end{align}
similar to the inverse see-saw one in \equ{iss} with the 
$\mu$-term as source of LNV.
Here the Majorana mass term for $N_{L}'$ arises after the spontaneous 
breaking of the electroweak symmetry (and the matter parity) with the
coefficient
\begin{align}
(\Lambda^{-1})_{cd}
=
2
\kappa_{cd}
\frac{\mu}{M^2_{\phi}}
\sim 
\mathcal{O}
\left(\frac{1}{\Lambda_{\rm NP}}\right)
\end{align}
suppressed by the new physics scale. 

The effective neutrino masses are then given by
\begin{align}
m_{\nu}
=
\frac{v_{u}^{3} v_{d}}{4}
Y_{\nu}^{\sf T}
(M^{-1})^{\sf T}
\Lambda^{-1}
M^{-1}
Y_{\nu}
 \sim
\mathcal{O}
\left(\frac{v^{4}}{\Lambda_{\rm NP}^{3}}\right)\,,
\end{align}
where $v_u=\sqrt{2}\langle H^0_u\rangle$ and $v_d=\sqrt{2}\langle H^0_d\rangle$. If we assume $\Lambda_{\mathrm{NP}} \sim 1 \, \text{TeV}$ and $m_{\nu} \sim 1 \, \text{ eV}$,
then we have the Dirac mass term for the $\overline{N_{R}}$-$\nu_{L}$ 
interaction with $Y_{\nu} v_{u} \sim 10 \, \text{ MeV}$,
which is smaller than in the ordinary see-saw scenario
but of the same order as the charged lepton masses, \ie,
$Y_{\nu}$ is not {\it extremely} small in comparison with
the other fermion Yukawa couplings.

In order to have the interactions in Eq~\eqref{equ:L-decon-3} and
to forbid the Majorana mass term for the SM singlet fermion 
$N_{R}$ and the Yukawa interaction with $N_{L}'$, 
we assign the following charges\footnote{Note that allowing the Yukawa
interaction in \equ{L-decon-3},  together with
Eq.~(\ref{eq:cond-dim7-1}) automatically gives the necessary conditions
to forbid these two terms.
} under a $\mathbb{Z}_5$:
\begin{gather}
 q_{H_{u}} = 0 \, , \quad q_{H_{d}} = 3 \, , \quad
q_{L} = 1 \, , \quad  q_{N_{R}} = q_{N_{L}'} = 1, \quad q_{\phi} = 3 \, .
\label{eq:charge-inverseseesaw-nonSUSY}
\end{gather}
Note that we cannot forbid the interaction $\overline{N_{R}^{c}} N_{R} \phi$ by any charge assignment,
which means that the (2,2) element ($\mu'$-term) in \equ{massmatrix-inverseseesaw} may actually be non-zero, but 
suppressed with respect to the Dirac masses of $N_{R}$ and
$N_{L}'$. Nevertheless, such a Majorana mass term $M_{R}$ for $N_{R}$ 
gives a contribution to neutrino masses proportional to 
$(v_u^4 v_d^2 M_{R})/(M^4\Lambda^2_{\trm{NP}})$, 
which is of second order, and can thus be omitted from this discussion.

It is interesting to compare our approach to the original inverse see-saw model. In the original model, the texture of the mass matrix is justified by the lepton number symmetry with the charge assignment $L(\nu_{L})=1$, $L(N_{R})=1$ and $L(N_{L}')=1$.  Then, the Majorana mass term of the $N_{L}'$ field (the $\mu$-term in \equ{iss}) is introduced by hand and its smallness is justified by the fact that it is the only lepton number violating quantity. In our model, the texture of the mass matrix is determined by the $\mathbb{Z}_5$ symmetry. Moreover, the Majorana mass for the $N_{L}'$ field is generated after electroweak symmetry (and matter parity) breaking and is naturally small since it is suppressed by the scale of new physics $\Lambda_{\trm{NP}}$. We thus implement what is sometimes called ``double see-saw'' rather  than inverse see-saw. Indeed we have one see-saw mechanism which generates a small Majorana mass for the new fermion singlet $N_{L}'$ suppressed by $\Lambda_{\mathrm{NP}}$, and another one which generates small neutrino masses suppressed by $M$.

In fact, this model, defined by the SM Lagrangian plus the interactions
displayed in Eq.~(\ref{equ:L-decon-3}), has more than a 
$\mathbb{Z}_5$ symmetry: 
it is also invariant under a new U(1) symmetry.\footnote{%
The new symmetry neither corresponds to lepton number nor hypercharge.
It contains $\mathbb{Z}_{5}$ which is often called 
``matter parity''.
With respect to the Higgs potential, it plays the same role as 
the Peccei-Quinn 
symmetry~\cite{Weinberg:1977ma,Wilczek:1977pj,Peccei:1977hh,Peccei:1977ur}.} 
This is potentially dangerous since the breaking of the electroweak 
symmetry also breaks this U(1) symmetry leading to a massless Goldstone
boson. 
However, this can be avoided by an enhanced scalar sector, provided the term
\begin{equation}
\frac{\lambda}{\Lambda_{\phi}}\phi^5\,,
\label{equ:phi5}
\end{equation}
appears in the effective Lagrangian after integration of the degrees of
freedom  of some high energy theory (here $\Lambda_{\phi}$ denotes the
typical scale).  We do not provide such a theory here explicitly since
it is not directly relevant for the generation of neutrino mass in this
context. 
Neverthless, we have checked that one can have an enhanced scalar sector
to produce Eq.~\eqref{equ:phi5} in this model, without having
massless Goldstone bosons and unwanted tadpole of additional scalars
at the same time.
Instead, we refer to \Sec~\ref{sec:higher} for an explicit model where neutrino masses depend crucially on the breaking of some U(1) symmetry down to $\mathbb{Z}_5$.

\begin{figure}[t!]
\unitlength=1cm
\begin{center}
\begin{picture}(5,3)
\put(0,0){\includegraphics[width=5cm]{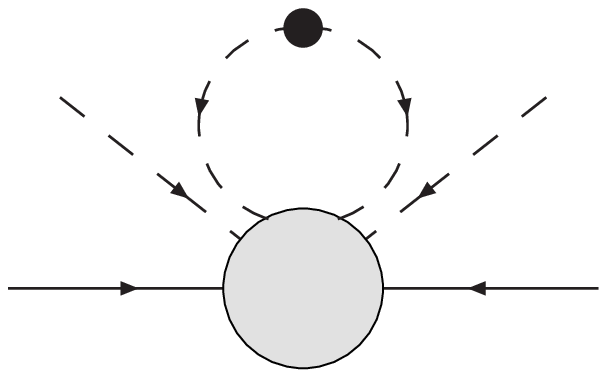}}
\put(0.2,0.8){$L$}
\put(4.6,0.8){$L$}
\put(0.3,2.3){$H_{u}$}
\put(4.4,2.3){$H_{u}$}
\put(1.3,2.5){$H_{d}$}
\put(3.15,2.5){$H_{u}$}
\put(2.3,3){$m_{3}^{2}$}
\end{picture}
\end{center}
\mycaption{One loop contribution with the soft breaking term of 
 Eq.~\eqref{eq:cross-term}.
 It is constructed by the soft breaking term and the dimension seven operator 
 which is allowed by the symmetry.
 Therefore, it is proportional to the dimension seven operator.
 The size of the operator is estimated as Eq.~(31), and
 it is expected to be always smaller than the dimension seven
 contribution by the loop suppression factor.
 }
\label{Fig:dim5-loop}
\end{figure}

As an alternative, one can introduce a soft violation of the U(1) symmetry (and also $\mathbb{Z}_{5}$)
\begin{equation}
\mathscr{V}_{\text{soft}} = m_{3}^{2} H_{d} {\rm i} \tau^{2} H_{u} +
 {\rm H.c.} \, ,
\label{eq:cross-term}
\end{equation}
where $m_{3}$ is assumed to be the electroweak scale.
This term is generally introduced in the THDM as the soft breaking
term of $\mathbb{Z}_{2}$ to forbid FCNC (Flavor Changing Neutral Current) processes~\cite{Glashow:1976nt,Barger:1989fj}.
The introduction of this soft term makes the Higgs phenomenology
MSSM-like. With this term, the Goldstone boson obtains the mass proportional
to $m_{3}$, which is identified with the CP odd Higgs boson, $A^{0}$, in
the THDM and the MSSM. This soft term also affects neutrino masses since
it explicitly violates $\mathbb{Z}_5$, which implies that the dimension
five Weinberg operator must appear at the loop level by insertion of
that term inside a loop.
Note that the loop diagram is constructed by closing the Higgs 
propagators of the dimension seven operator $LLH_{u} H_{u} H_{d} H_{u}$. 
Therefore, it has to be proportional to the 
dimension seven contribution (\cf, Fig.~\ref{Fig:dim5-loop}).
The size of the contribution is estimated as
\begin{equation}
\frac{1}{16\pi^{2}} \frac{m_{3}^{2}}{\Lambda_{\rm NP}^{3}} (LLH_{u}H_{u}),
\end{equation}
which is still suppressed with respect to the tree level dimension seven contribution
by the loop suppression factor. Therefore, the introduction of a
soft term would not disturb our main line of argumentation.\footnote{Note that compared to the
loop contributions from closing the $H^\dagger H$ loop, there is factor
of $m_3^2$ in the numerator compared to $\Lambda_{\mathrm{NP}}^2$, which
means that this contribution is effectively suppressed as strong as the
one loop $d=7$ operator instead of the one loop $d=5$
operator. Therefore, our usual argumentation (in the introduction) on
closing the loops does not apply.}

\subsection{Decomposition (b): The $\boldsymbol{\epsilon}$-term}

For the decomposition (b) in \figu{decomp}, we introduce two chiral
fermions, singlets of the SM model $N_{R}$ (right-handed) and  
$N_{L}'$ (left-handed), and a SU(2)$_{L}$ doublet scalar $\Phi$. We then need the following relevant interactions:
\begin{align}
 \mathscr{L}
 =& 
 \mathscr{L}_{\rm SM}
 +
 \Bigl[
 (Y_{\nu})_{a\alpha}\,
 (\overline{N_{R}})_{a}
 H_{u} 
 {\rm i} \tau^{2} 
 L_{\alpha}
 +
 (Y_{\nu}')_{a\alpha}\,
 (\overline{N_{L}'^{c}})_{a}
 \Phi^{\dagger} L_{\alpha}
 +
 \zeta
 \{
 (H_{d} {\rm i} \tau^{2} H_{u}) 
 (\Phi {\rm i} \tau^{2} H_{u})
 \}
 \nonumber 
 \\
 &\hspace{1.5cm}
 +
 (\overline{N_{R}})_{a} M_{ab} (N_{L}')_{b}
 +
 {\rm H.c.}
 \Bigr]
 +
 M_{\Phi}^{2}
 \Phi^{\dagger} \Phi
 +
 \cdots.
\end{align}
These lead to the mass matrix
\begin{align}
\mathscr{L}=
\frac{1}{2}
\begin{pmatrix}
\overline{\nu_{L}^{c}}_{\alpha}
&
\overline{N_{R}}_{a}
&
(\overline{N_{L}'^{c}})_{c}
\end{pmatrix}
\begin{pmatrix}
 0 
 &(Y_{\nu}^{\sf T})_{\alpha b} \langle H_{u}^{0} \rangle 
 & 
 \frac{\zeta (Y_{\nu}'^{\sf T})_{\alpha d}}{M_{\Phi}^{2}}
 \langle H_{d}^{0} \rangle \langle H_{u}^{0} \rangle^{2}
 \\
 (Y_{\nu})_{a \beta} \langle H_{u}^{0} \rangle 
& 0 
& M_{ad}
 \\
 \frac{\zeta (Y_{\nu}')_{c \beta }}{M_{\Phi}^{2}}
 \langle H_{d}^{0} \rangle \langle H_{u}^{0} \rangle^{2} 
 & 
 (M^{\sf T})_{cb}
 & 
 0
\end{pmatrix}
\begin{pmatrix}
\nu_{L\beta} \\ (N_{R}^{c})_{b} \\ N_{L d}'
\end{pmatrix}
+{\rm H.c.}.
\label{equ:massmatrix-inverseseesaw-Nr2}
\end{align}
with a non-trivial $(1,3)$ element ($\epsilon$-term in \equ{iss}) 
again suppressed by the new physics scale. 
The joint presence of the three entries violates lepton number 
and yields the neutrino mass
\be
m_\nu = \frac{\zeta v_u^3v_d}{4M_{\Phi}^2}
\left(
Y_\nu^{\sf T} M^{-1} Y_{\nu}' +
Y_{\nu}'^{\sf T} (M^{-1})^{\sf T} Y_{\nu}
\right)  
\sim \mathcal{O}\left( \frac{v^4}{\Lambda_{\rm NP}^3} \right) \,.
\ee
A contribution of the same order of that of the case we considered previously.

The conditions on the charges imposed by the fundamental interactions can be implemented by the following assignments
under a $\mathbb{Z}_5$:
\begin{gather}
q_{H_{u}} = 0 \, , \quad q_{H_{d}} = 3, \, \quad q_{L} = 1 \, , \quad 
q_{N_{R}} = q_{N_{L}'} = 1, \quad q_{\Phi} = 2 \, .
\label{equ:charge-inverseseesaw-epsilon-nonSUSY}
\end{gather}
This model is also a double see-saw but involves the product of two
Dirac masses, contrary to the previous case where Majorana masses were
involved. Again, the U(1) has to be broken explicitly.


\section{Generalization of standard see-saws}
\label{sec:sys}


\begin{figure}[tb]
\begin{center}
\unitlength=1cm
\begin{picture}(16,3.5)
\put(0,0){\includegraphics[width=4cm]{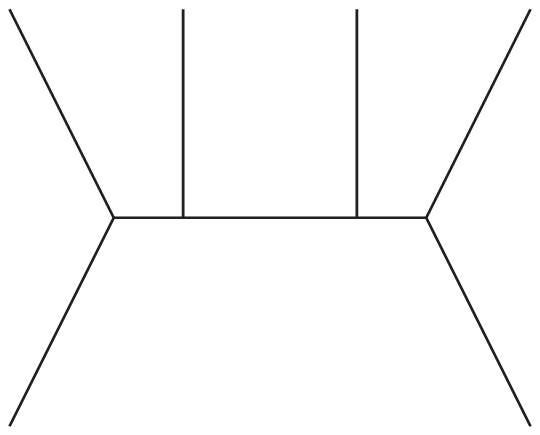}}
\put(4,0){\includegraphics[width=4cm]{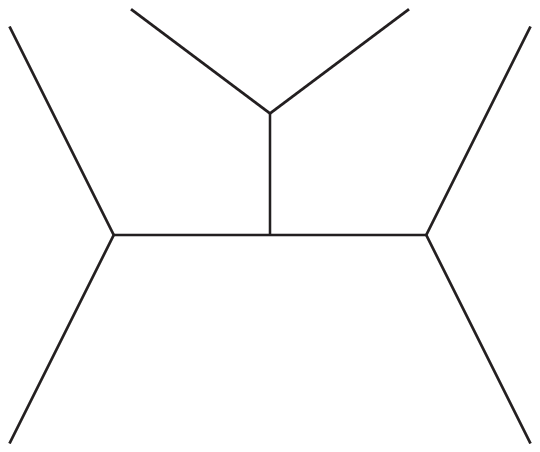}}
\put(8,0){\includegraphics[width=4cm]{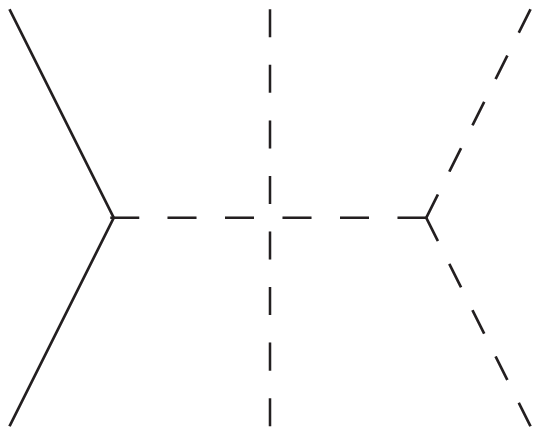}}
\put(12,0){\includegraphics[width=4cm]{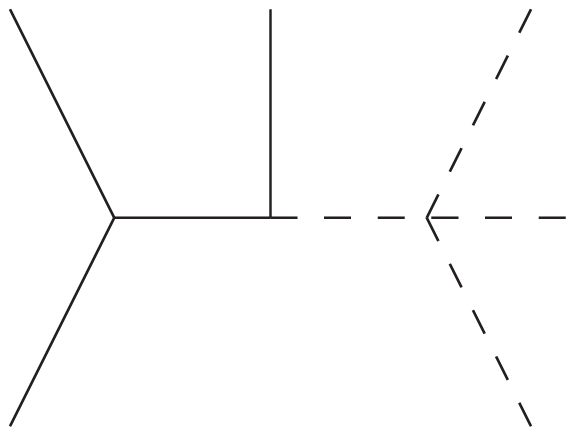}}
\put(1.2,-0.2){Topology 1}
\put(5.2,-0.2){Topology 2}
\put(9.2,-0.2){Topology 3}
\put(13.2,-0.2){Topology 4}
\end{picture}
\mycaption{\label{fig:top}Possible topologies for the tree level decomposition of the dimension seven operator 
$LLH_{u} H_{u} H_{d} H_{u}$.
The dashed lines denote always scalars (scalar mediators or the Higgs 
doublet). The solid lines in Topology~1, 2, and 4 should be interpreted 
as fermions or scalars depending on the decomposition.}
\end{center}
\end{figure}

Here we show all possible decompositions of the dimension seven 
 operator $LL H_{u} H_{u} H_{d} H_{u}$ (\#4 in \Tab~\ref{tab:operators}) at tree level.
We do not go into the details of the models, such as the Lagrangian and the
matter parity conditions to implement it. Therefore, these results should
be interpreted as the necessary conditions to find tree level neutrino
mass models from this dimension seven operator. For the new fields, we follow the notation 
in \Ref~\cite{Gavela:2008ra}. They are denoted by  ${\bf X}^{\mathcal{L}}_Y$, where
\begin{itemize}
\item
 ${\bf X}$ denotes the SU(2) nature, \ie, singlet ${\bf 1}$, doublet ${\bf 2}$, or triplet ${\bf 3}$.
\item
     $\mathcal{L}$ refers to the Lorentz nature, \ie, scalar ($s$), vector
     ($v$),  left-handed ($L$) or right-handed ($R$) chiral fermion.
\item
 $Y$ refers to the hypercharge $Y=Q-I^W_3$.
\end{itemize}
The possible topologies of the Feynman diagrams are shown in \figu{top}.

\begin{table}[p]
\footnotesize{
\begin{tabular}{ccccccc}
\hline \hline
&&&& \multicolumn{3}{c}{Phenom.}
\\
\# & Operator & Top. & Mediators & NU & $\delta g_{L}$ & 4$\ell$  
\\
\hline
1
&
     $(H_{u} {\rm i} \tau^{2} \overline{L^{c}})
     (H_{u} {\rm i} \tau^{2} L)(H_{d} {\rm i}
     \tau^{2} H_{u})$
&
2
&
${\bf 1}^{R}_{0}$, ${\bf 1}^{L}_{0}$, ${\bf 1}^{s}_{0}$
&
$\checkmark$
&
&
\\
2
&
     $(H_{u} {\rm i} \tau^{2} \vec{\tau} \overline{L^{c}})
     (H_{u} {\rm i} \tau^{2} L)
     (H_{d} {\rm i} \tau^{2} \vec{\tau} H_{u})$
&
2
&
${\bf 3}^{R}_{0}$, ${\bf 3}^{L}_{0}$
${\bf 1}^{R}_{0}$, ${\bf 1}^{L}_{0}$,
${\bf 3}^{s}_{0}$
&
$\checkmark$
&
$\checkmark$
&
\\
3
&
     $(H_{u} {\rm i} \tau^{2} \vec{\tau} \overline{L^{c}})
     (H_{u} {\rm i} \tau^{2} \vec{\tau} L)
     (H_{d} {\rm i} \tau^{2} H_{u})$
&
2
&
${\bf 3}^{R}_{0}$, ${\bf 3}^{L}_{0}$, ${\bf 1}^{s}_{0}$
&
$\checkmark$
&
$\checkmark$
&
\\
4
&
$ 
     (-{\rm i} \epsilon^{abc})
     (H_{u} {\rm i} \tau^{2} \tau^{a} \overline{L^{c}})
     (H_{u} {\rm i} \tau^{2} \tau^{b} L)
     (H_{d} {\rm i} \tau^{2} \tau^{c} H_{u})$
&
2
&
${\bf 3}^{R}_{0}$, ${\bf 3}^{L}_{0}$, ${\bf 3}^{s}_{0}$
&
$\checkmark$
&
$\checkmark$
&
\\
5 
&
     $(\overline{L^{c}} {\rm i} \tau^{2} \vec{\tau} L) 
     (H_{d} {\rm i} \tau^{2} H_{u}) 
     (H_{u} {\rm i} \tau^{2} \vec{\tau}  H_{u})$
&
2/3
&
${\bf 3}^{s}_{-1}$, ${\bf 3}^{s}_{-1}$/${\bf 1}^{s}_{0}$
&
&
&
$\checkmark$
 \\
\hline
6 
&
$
     (-{\rm i} \epsilon^{abc})
     (\overline{L^{c}} {\rm i} \tau^{2} \tau^{a} L) 
     (H_{d} {\rm i} \tau^{2} \tau^{b} H_{u}) 
     (H_{u} {\rm i} \tau^{2} \tau^{c}  H_{u})$
&
2/3
&
${\bf 3}^{s}_{-1}$,
${\bf 3}^{s}_{-1}$/${\bf 3}^{s}_{0}$
&
&
&
$\checkmark$
\\
7
&
$(H_{u} {\rm i} \tau^{2} \overline{L^{c}})
(L {\rm i} \tau^{2} \vec{\tau} H_{d})
(H_{u} {\rm i} \tau^{2} \vec{\tau} H_{u})$
&
2
&
${\bf 1}^{R}_{0}$, ${\bf 1}^{L}_{0}$,
${\bf 3}^{R}_{0}$, ${\bf 3}^{L}_{0}$,
${\bf 3}^{s}_{-1}$
&
$\checkmark$
&
$\checkmark$
&
\\
8
&
$(-{\rm i} \epsilon^{abc})
(H_{u} {\rm i} \tau^{2} \tau^{a} \overline{L^{c}})
(L {\rm i} \tau^{2} \tau^{b} H_{d})
(H_{u} {\rm i} \tau^{2} \tau^{c} H_{u})$
&
2
&
${\bf 3}^{R}_{0}$, ${\bf 3}^{L}_{0}$,
${\bf 3}^{R}_{0}$, ${\bf 3}^{L}_{0}$,
${\bf 3}^{s}_{-1}$
&
$\checkmark$
&
$\checkmark$
&
\\
9
&
$(H_{u} {\rm i} \tau^{2} \overline{L^{c}})
 ({\rm i} \tau^{2} H_{u})
 (L)
 (H_{d} {\rm i} \tau^{2} H_{u})$
&
1
&
${\bf 1}^{R}_{0}$, ${\bf 1}^{L}_{0}$, 
${\bf 2}^{R}_{-1/2}$, ${\bf 2}^{L}_{-1/2}$,
${\bf 1}^{s}_{0}$
&
$\checkmark$
&
&
\\
10
&
$(H_{u} {\rm i} \tau^{2} \vec{\tau} \overline{L^{c}})
 ({\rm i} \tau^{2} \vec{\tau} H_{u})
 (L)
 (H_{d} {\rm i} \tau^{2} H_{u})$
&
1
&
${\bf 3}^{R}_{0}$, ${\bf 3}^{L}_{0}$, 
${\bf 2}^{R}_{-1/2}$, ${\bf 2}^{L}_{-1/2}$,
${\bf 1}^{s}_{0}$
&
$\checkmark$
&
$\checkmark$
&
\\
\hline
11
&
$(H_{u} {\rm i} \tau^{2} \overline{L^{c}})
 ({\rm i} \tau^{2} H_{u})
 (\vec{\tau} L)
 (H_{d} {\rm i} \tau^{2} \vec{\tau} H_{u})$
&
1
&
${\bf 1}^{R}_{0}$, ${\bf 1}^{L}_{0}$, 
${\bf 2}^{R}_{-1/2}$, ${\bf 2}^{L}_{-1/2}$,
${\bf 3}^{s}_{0}$
&
$\checkmark$
&
&
\\
12
&
$(H_{u} {\rm i} \tau^{2} \tau^{a} \overline{L^{c}})
 ({\rm i} \tau^{2} \tau^{a} H_{u})
 (\tau^{b} L)
 (H_{d} {\rm i} \tau^{2} \tau^{b} H_{u})$
&
1
&
${\bf 3}^{R}_{0}$, ${\bf 3}^{L}_{0}$, 
${\bf 2}^{R}_{-1/2}$, ${\bf 2}^{L}_{-1/2}$,
${\bf 3}^{s}_{0}$
&
$\checkmark$
&
$\checkmark$
&
\\
13
&
     $(H_{u} {\rm i} \tau^{2} \overline{L^{c}})(L) 
     ({\rm i} \tau^{2} H_{u})
     (H_{d} {\rm i} \tau^{2} H_{u})$
&
1/4
&
${\bf 1}^{R}_{0}$, ${\bf 1}^{L}_{0}$,
${\bf 2}^{s}_{-1/2}$, (${\bf 1}^{s}_{0}$)
&
$\checkmark$
&
&
\\
14
&
     $(H_{u} {\rm i} \tau^{2} \vec{\tau} \overline{L^{c}})
     (\vec{\tau} L) 
     ({\rm i} \tau^{2} H_{u})
     (H_{d} {\rm i} \tau^{2} H_{u})$
&
1/4
&
${\bf 3}^{R}_{0}$, ${\bf 3}^{L}_{0}$,
${\bf 2}^{s}_{-1/2}$, (${\bf 1}^{s}_{0}$)
&
$\checkmark$
&
$\checkmark$
&
\\
15
&
     $(H_{u} {\rm i} \tau^{2} \overline{L^{c}})(L) 
     ({\rm i} \tau^{2} \vec{\tau} H_{u})
     (H_{d} {\rm i} \tau^{2} \vec{\tau} H_{u})$
&
1/4
&
${\bf 1}^{R}_{0}$, ${\bf 1}^{L}_{0}$,
${\bf 2}^{s}_{-1/2}$, (${\bf 3}^{s}_{0}$)
&
$\checkmark$
&
&
\\
\hline
16
&
     $(H_{u} {\rm i} \tau^{2} \tau^{a} \overline{L^{c}})( \tau^{a} L) 
     ({\rm i} \tau^{2} \tau^{b} H_{u})
     (H_{d} {\rm i} \tau^{2} \tau^{b} H_{u})$
&
1/4
&
${\bf 3}^{R}_{0}$, ${\bf 3}^{L}_{0}$,
${\bf 2}^{s}_{-1/2}$, (${\bf 3}^{s}_{0}$)
&
$\checkmark$
&
$\checkmark$
&
\\
17
&
     $(H_{u} {\rm i} \tau^{2} \overline{L^{c}})
     (H_{d}) ({\rm i} \tau^{2} H_{u}) 
     (H_{u} {\rm i} \tau^{2} L)$
&
1
&
${\bf 1}^{R}_{0}$, ${\bf 1}^{L}_{0}$,
${\bf 2}^{R}_{-1/2}$, ${\bf 2}^{L}_{-1/2}$
&
$\checkmark$
&
&
\\
18
&
     $(H_{u} {\rm i} \tau^{2}\vec{\tau} \overline{L^{c}})
     (\vec{\tau} H_{d}) 
     ({\rm i} \tau^{2} H_{u}) 
     (H_{u} {\rm i} \tau^{2} L)$
&
1
&
${\bf 3}^{R}_{0}$, ${\bf 3}^{L}_{0}$,
${\bf 2}^{R}_{-1/2}$, ${\bf 2}^{L}_{-1/2}$,
${\bf 1}^{R}_{0}$, ${\bf 1}^{L}_{0}$
&
$\checkmark$
&
$\checkmark$
&
\\
19
&
     $(H_{u} {\rm i} \tau^{2} \overline{L^{c}})
     (H_{d}) ({\rm i} \tau^{2} \vec{\tau} H_{u}) 
     (H_{u} {\rm i} \tau^{2} \vec{\tau} L)$
&
1
&
${\bf 1}^{R}_{0}$, ${\bf 1}^{L}_{0}$,
${\bf 2}^{R}_{-1/2}$, ${\bf 2}^{L}_{-1/2}$,
${\bf 3}^{R}_{0}$, ${\bf 3}^{L}_{0}$
&
$\checkmark$
&
$\checkmark$
&
\\
20
&
     $(H_{u} {\rm i} \tau^{2} \tau^{a} \overline{L^{c}})
     (\tau^{a} H_{d}) ({\rm i} \tau^{2} \tau^{b} H_{u}) 
     (H_{u} {\rm i} \tau^{2} \tau^{b} L)$
&
1
&
${\bf 3}^{R}_{0}$, ${\bf 3}^{L}_{0}$,
${\bf 2}^{R}_{-1/2}$, ${\bf 2}^{L}_{-1/2}$,
&
$\checkmark$
&
$\checkmark$
&
\\
\hline
21
&
$(\overline{L^{c}} {\rm i} \tau^{2} \tau^{a} L) 
(H_{u} {\rm i} \tau^{2} \tau^{a})
(\tau^{b} H_{d})
(H_{u} {\rm i} \tau^{2} \tau^{b} H_{u})$
&
1/4
&
${\bf 3}^{s}_{-1}$, ${\bf 2}^{s}_{+1/2}$, (${\bf 3}^{s}_{-1}$)
&
&
&
$\checkmark$
\\
22
&
$(\overline{L^{c}} {\rm i} \tau^{2} \tau^{a} L) 
(H_{d} {\rm i} \tau^{2} \tau^{a})
(\tau^{b} H_{u})
(H_{u} {\rm i} \tau^{2} \tau^{b} H_{u})$
&
1/4
&
${\bf 3}^{s}_{-1}$, ${\bf 2}^{s}_{+3/2}$, (${\bf 3}^{s}_{-1}$)
&
&
&
$\checkmark$
\\
23
&
$(\overline{L^{c}} {\rm i} \tau^{2} \vec{\tau} L) 
(H_{u} {\rm i} \tau^{2} \vec{\tau})
(H_{u})
(H_{d} {\rm i} \tau^{2} H_{u})$
&
1/4
&
${\bf 3}^{s}_{-1}$, ${\bf 2}^{s}_{+1/2}$, (${\bf 1}^{s}_{0}$)
&
&
&
$\checkmark$
\\
24
&
$(\overline{L^{c}} {\rm i} \tau^{2} \tau^{a} L) 
(H_{u} {\rm i} \tau^{2} \tau^{a})
(\tau^{b} H_{u})
(H_{d} {\rm i} \tau^{2} \tau^{b} H_{u})$
&
1/4
&
${\bf 3}^{s}_{-1}$, ${\bf 2}^{s}_{+1/2}$, (${\bf 3}^{s}_{0}$)
&
&
&
$\checkmark$
\\
25
&
$(H_{d} {\rm i} \tau^{2} H_{u}) 
     (\overline{L^{c}} {\rm i} \tau^{2})
     (\vec{\tau} L)
     (H_{u} {\rm i} \tau^{2} \vec{\tau} H_{u})$
&
1
&
${\bf 1}^{s}_{0}$, 
${\bf 2}^{L}_{+1/2}$, ${\bf 2}^{R}_{+1/2}$,
${\bf 3}^{s}_{-1}$
\\
\hline
26
&
$(H_{d} {\rm i} \tau^{2} \tau^{a} H_{u}) 
     (\overline{L^{c}} {\rm i} \tau^{2} \tau^{a})
     (\tau^{b} L)
     (H_{u} {\rm i} \tau^{2} \tau^{b} H_{u})$
&
1
&
${\bf 3}^{s}_{0}$, 
${\bf 2}^{L}_{+1/2}$, ${\bf 2}^{R}_{+1/2}$,
${\bf 3}^{s}_{-1}$
\\
27
&
$(H_{u} {\rm i} \tau^{2} \overline{L^{c}})
({\rm i} \tau^{2} H_{d})
(\vec{\tau} L)
(H_{u} {\rm i} \tau^{2} \vec{\tau} H_{u})$
&
1
&
${\bf 1}^{R}_{0}$, ${\bf 1}^{L}_{0}$,
${\bf 2}^{R}_{+1/2}$, ${\bf 2}^{L}_{+1/2}$,
${\bf 3}^{s}_{-1}$
&
$\checkmark$
&
&
\\
28
&
$(H_{u} {\rm i} \tau^{2} \tau^{a} \overline{L^{c}})
({\rm i} \tau^{2} \tau^{a} H_{d})
(\tau^{b} L)
(H_{u} {\rm i} \tau^{2} \tau^{b} H_{u})$
&
1
&
${\bf 3}^{R}_{0}$, ${\bf 3}^{L}_{0}$,
${\bf 2}^{R}_{+1/2}$, ${\bf 2}^{L}_{+1/2}$,
${\bf 3}^{s}_{-1}$
&
$\checkmark$
&
$\checkmark$
&
\\
29
&
$(H_{u} {\rm i} \tau^{2} \overline{L^{c}})
(L)
({\rm i} \tau^{2} \vec{\tau} H_{d})
(H_{u} {\rm i} \tau^{2} \vec{\tau} H_{u})$
&
1/4
&
${\bf 1}^{R}_{0}$, ${\bf 1}^{L}_{0}$,
${\bf 2}^{s}_{+1/2}$, 
(${\bf 3}^{s}_{-1}$)
&
$\checkmark$
&
&
\\
30
&
$(H_{u} {\rm i} \tau^{2} \tau^{a} \overline{L^{c}})
(\tau^{a} L)
({\rm i} \tau^{2} \tau^{b} H_{d})
(H_{u} {\rm i} \tau^{2} \tau^{b} H_{u})$
&
1/4
&
${\bf 3}^{R}_{0}$, ${\bf 3}^{L}_{0}$,
${\bf 2}^{s}_{+1/2}$,
(${\bf 3}^{s}_{-1}$)
&
$\checkmark$
&
$\checkmark$
&
\\
\hline
31
&
$(\overline{L^{c}} {\rm i} \tau^{2} \tau^{a} H_{d})
     ({\rm i} \tau^{2} \tau^{a} H_{u})
     (\tau^{b} L)
     (H_{u} {\rm i} \tau^{2} \tau^{b} H_{u})$
&
1
&
${\bf 3}^{L}_{+1}$, ${\bf 3}^{R}_{+1}$,
${\bf 2}^{L}_{+1/2}$, ${\bf 2}^{R}_{+1/2}$,
${\bf 3}^{s}_{-1}$
&
$\checkmark$
&
$\checkmark$
&
\\
32
&
$(\overline{L^{c}} {\rm i} \tau^{2} \tau^{a} H_{d})
     (\tau^{a} L)
     ({\rm i} \tau^{2} \tau^{b} H_{u})
     (H_{u} {\rm i} \tau^{2} \tau^{b} H_{u})$
&
1/4
&
${\bf 3}^{L}_{+1}$, ${\bf 3}^{R}_{+1}$,
${\bf 2}^{s}_{-3/2}$,
(${\bf 3}^{s}_{-1}$)
&
$\checkmark$
&
$\checkmark$
&
\\
33
&
$(\overline{L^{c}} {\rm i} \tau^{2} \vec{\tau} H_{d})
({\rm i} \tau^{2} \vec{\tau} H_{u})
(H_{u})
(H_{u} {\rm i} \tau^{2} L)$
&
1
&
${\bf 3}^{L}_{+1}$, ${\bf 3}^{R}_{+1}$,
${\bf 2}^{L}_{-3/2}$, ${\bf 2}^{R}_{-3/2}$,
${\bf 1}^{L}_{0}$, ${\bf 1}^{R}_{0}$
&
$\checkmark$
&
$\checkmark$
&
\\
34
&
$(\overline{L^{c}} {\rm i} \tau^{2} \tau^{a} H_{d})
({\rm i} \tau^{2} \tau^{a} H_{u})
(\tau^{b} H_{u})
(H_{u} {\rm i} \tau^{2} \tau^{b} L)$
&
1
&
${\bf 3}^{L}_{+1}$, ${\bf 3}^{R}_{+1}$,
${\bf 2}^{L}_{-3/2}$, ${\bf 2}^{R}_{-3/2}$,
${\bf 3}^{L}_{0}$, ${\bf 3}^{R}_{0}$
&
$\checkmark$
&
$\checkmark$
&
\\
\hline \hline
\end{tabular}
} 
\mycaption{\label{tab:decomp}
Decompositions of the effective dimension seven operator
$LLH_{u} H_{u} H_{d} H_{u}$. The brackets in the operators show the fundamental interactions, \ie, each operator corresponds to a Feynman diagram with the topology listed in the 3rd column (\cf, \figu{top}). The fourth column shows the SM quantum numbers of the required mediators, where each symbol represents a separate new field. The abbreviation ``$X$/$Y$'' means that either $X$ or $Y$ or both are different possibilities, depending on the topology, whereas the abbreviation ``($X$)'' means that $X$ is optional, depending on the topology. The last columns indicate the phenomenology one may expect in this model, where ``NU'' stands for non-unitarity of the lepton mixing matrix, ``$\delta g_L$'' for a shift of the neutral current coupling with charged leptons, and ``$4 \ell$'' for charged lepton flavor violation or non-standard neutrino interactions. }
\end{table}

We present our results in \Tab~\ref{tab:decomp}. In this table, we show the 
decompositions of all possible combinations leading to the effective operator $LLH_{u} H_{u} H_{d} H_{u}$.

 The decompositions which we have discussed in \Sec~\ref{sec:invdecomp} are \#1 and \#13 in the table. The table is useful to read off the potential models for any possible set of mediators. For example, one can read off from \Tab~\ref{tab:decomp} the generic realizations\footnote{By generic realization we mean a decomposition which includes one or two copies of the original mediator (${\bf 1}^{R/L}_{0}$ for the Type I see-saw, ${\bf 3}^{s}_{-1}$ for the Type II see-saw and ${\bf 3}^{R/L}_{0}$ for the Type III see-saw) plus extra scalar mediators.} of the standard type~I,~II, and~III see-saw mechanisms by using their field content and additional mediators:
\begin{description}
\item[Type~I (fermionic singlet mediator)] Operators \#1, \#13, \#15, and \#29 are simple generalizations.
In fact, our decompositions in \Sec~\ref{sec:invdecomp} represent some of the simplest possible generalizations of the
type~I see-saw mechanism, which require only three types of new fields in total. 
\item[Type~II (scalar triplet mediator)] Operators \#5, \#6, \#21, \#22, \#23, and \#24 are the simplest possibilities.
For example, \#5 requires an additional triplet scalar and/or singlet scalar. 
\item[Type~III (fermionic triplet mediator)] Operators \#3, \#4, \#14, \#16, \#30 are possible options. For example, operator \#3 is the natural type~III counterpart of the inverse see-saw mechanism in the previous section.
\end{description}
One can also reads from  \Tab~\ref{tab:decomp} that some decompositions
are combinations of different types of see-saws: for example operator
\#2 can be viewed as a Type I + Type III see-saw.
Note that 
the decomposition shown in Ref.\cite{Babu:2009aq}
does not appear in \Tab~\ref{tab:decomp} 
because we concentrate on the decomposition 
with SU(2)$_{L}$ singlet, doublet, and triplet mediators.

\section{Additional suppression mechanisms}
\label{sec:higher}

In this section, we qualitatively sketch options for additional suppression 
mechanisms compared to the simplest possibility, the tree level decompositions of the
$d=7$ operators. For this discussion, it is useful to consider the following
expansion of the effective operators, where $\mathscr{L}_{d=D}^{(k)}$ stands
for the dimension $D$ contribution to the $k$-loop correction:
\begin{eqnarray}
\mathscr{L}_{\text{eff}}
=
\mathscr{L}_{\rm SM}
&
+ &
\frac{1}{\Lambda_{\rm NP}} 
\left[
\mathscr{L}_{d=5}^{(0)}
+
\delta
\mathscr{L}_{d=5}^{(1)}
+
\delta
\mathscr{L}_{d=5}^{(2)}
+
\cdots
\right]
\nonumber
\\
&
+ &
\frac{1}{\Lambda_{\rm NP}^{3}} 
\left[
\mathscr{L}_{d=7}^{(0)}
+
\delta
\mathscr{L}_{d=7}^{(1)}
+
\delta
\mathscr{L}_{d=7}^{(2)}
+
\cdots
\right]
\nonumber
\\
&
+ &
\frac{1}{\Lambda_{\rm NP}^{5}} 
\left[
\mathscr{L}_{d=9}^{(0)}
+
\delta
\mathscr{L}_{d=9}^{(1)}
+
\delta
\mathscr{L}_{d=9}^{(2)}
+
\cdots
\right]
\nonumber
\\
&
+ &
\frac{1}{\Lambda_{\rm NP}^{7}} 
\left[
\mathscr{L}_{d=11}^{(0)}
+
\delta
\mathscr{L}_{d=11}^{(1)}
+
\delta
\mathscr{L}_{d=11}^{(2)}
+
\cdots
\right]
\nonumber
\\
&+ & \cdots
\label{equ:Leffgen}
\end{eqnarray}
In general, the vertical expansion is controlled by the new symmetry, whereas the horizontal expansion is suppressed by the loop suppression factor. If, for instance, we go to $d=7$, we can switch off the first row in \equ{Leffgen} by imposing a new U(1) symmetry, and there is no need to worry about loop contributions at $d=5$. In this case, $\mathscr{L}_{d=7}^{(0)}$ gives the leading contribution for neutrino mass generation, as it was used the previous two sections.

 However, if one wants to implement additional suppression from higher dimensional operators either in the horizontal (loops) or vertical (higher $d$) direction, it is necessary to ensure that the discussed contribution is the leading order effect. For instance, it is {\em a priori} not clear which of the operators 
$\delta \mathscr{L}_{d=7}^{(1)}$ and $\mathscr{L}_{d=9}^{(0)}$ gives a larger contribution if both are allowed. As already discussed in the introduction, if $1/(16 \pi^2) \gtrsim (v/\Lambda_{\mathrm{NP}})^2$, or
$\Lambda_{\mathrm{NP}} \gtrsim 3 \, \mathrm{TeV}$, one would generically expect that the loop contributions are larger than the ones from the higher dimensional operators.  However, a too low new physics scale $\sim \mathrm{TeV}$
may be potentially harmful for a loop model if there are higher
dimensional tree level contributions.  
We will discuss a two loop model from $\delta \mathscr{L}_{d=7}^{(2)}$
in \Sec~\ref{sec:loop}. Since there is no contribution from
$\mathscr{L}_{d=7}^{(0)}$, $\delta \mathscr{L}_{d=7}^{(1)}$, and
$\mathscr{L}_{d=9}^{(0)}$ in this model, it gives the leading
contribution to neutrino mass for $\Lambda_{\mathrm{NP}} \gtrsim 3 \,
\mathrm{TeV}$. If $\Lambda_{\mathrm{NP}} \lesssim 3 \, \mathrm{TeV}$, 
$\mathscr{L}_{d=11}^{(0)}$ and $\delta \mathscr{L}_{d=9}^{(1)}$ 
have to be avoided as well. 
Furthermore, we show an example for $d=9$ from $\mathscr{L}_{d=9}^{(0)}$ 
in \Sec~\ref{sec:d9}.

\subsection{Higher than $\boldsymbol{d=7}$ at tree level}
\label{sec:d9}

\begin{figure}[tb]
\begin{center}
\unitlength=1cm
\begin{picture}(6,6)
\put(0,0){\includegraphics[width=6cm]{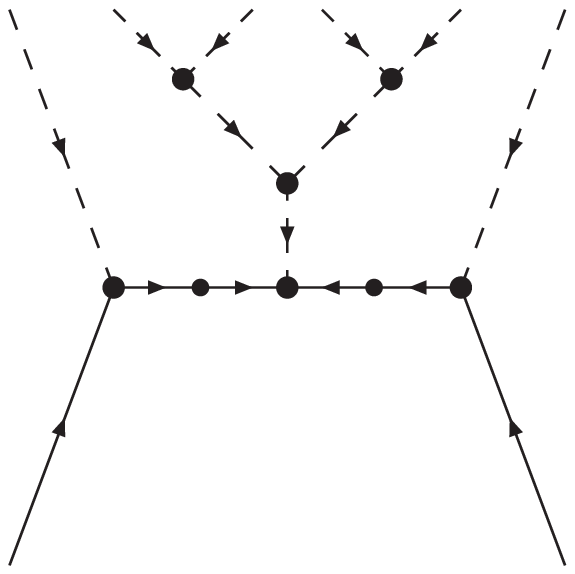}}
\put(0,0){$L$}
\put(5.8,0){$L$}
\put(0,5.8){$H_{u}$}
\put(5.7,5.8){$H_{u}$}
\put(1,5.7){$H_{d}$}
\put(2.5,5.7){$H_{u}$}
\put(3.2,5.7){$H_{d}$}
\put(4.6,5.7){$H_{u}$}
\put(2.6,3.2){$\phi$}
\put(2.1,4.2){$\varphi$}
\put(3.7,4.2){$\varphi$}
\put(1.5,2.4){$N_{R}$}
\put(2.4,2.4){$N_{L}'$}
\put(4.1,2.4){$N_{R}$}
\put(3.3,2.4){$N_{L}'$}
\put(3,-0.9){(c)}
\end{picture}
\hspace{1cm}
\begin{picture}(7,6)
\put(0,0.3){\includegraphics[width=7cm]{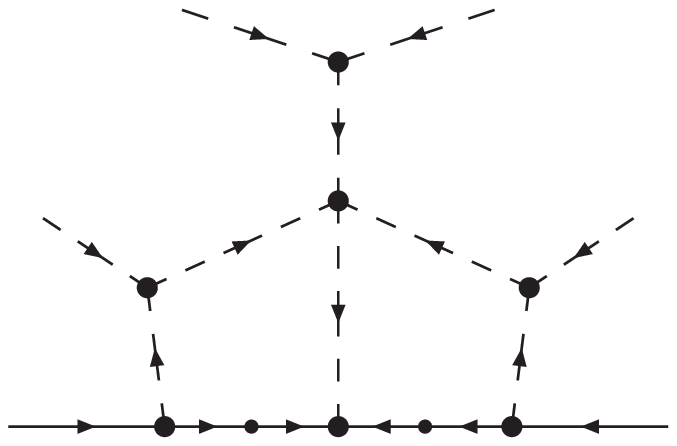}}
\put(0,0){$L$}
\put(6.7,0){$L$}
\put(0,2.6){$H_{u}$}
\put(6.5,2.6){$H_{u}$}
\put(1.5,4.7){$H_{d}$}
\put(5,4.7){$H_{u}$}
\put(1.1,1){$H_{d}$}
\put(5.4,1){$H_{d}$}
\put(3,3.2){$S$}
\put(2.2,2.4){$S$}
\put(4.4,2.4){$S$}
\put(3.7,1.2){$\phi$}
\put(1.9,0){$N_{R}$}
\put(2.7,0){$N_{L}'$}
\put(4.5,0){$N_{R}$}
\put(3.7,0){$N_{L}'$}
\put(3.5,-0.9){(d)}
\end{picture}
\vspace*{1cm}
\mycaption{Left panel: A possible tree level decomposition of the dimension nine operator 
$LLH_{u}H_{u}H_{d}H_{u}H_{d}H_{u}$ for the generation of neutrino mass.
 It  can also be interpreted within  the inverse see-saw framework. Right panel: A possible
two loop decomposition of the dimension seven operator $LLH_{u} H_{u} H_{d} H_{u}$.
}
\label{fig:dec-dim9}
\end{center}
\end{figure}

Here we qualitatively sketch an example of neutrino mass
generation from $\mathscr{L}_{d=9}^{(0)}$
($d=9$, tree level). The relevant diagram, corresponding to \#12
in \Tab~\ref{tab:operators}, is shown in \figu{dec-dim9} (c).
We introduce two SM singlet chiral fermions $N_{R}$ (right-handed) and 
$N_{L}'$ (left-handed), and two SM singlet scalars $\phi$ and $\varphi$.
The interaction Lagrangian is given by
\begin{align}
\mathscr{L}
 =&
 \mathscr{L}_{\rm SM}
 +
 \Bigl[
 (Y_{\nu})_{a \alpha}
 (\overline{N_{R}})_{a}
 H_{u} 
 {\rm i} \tau^{2} 
 L_{\alpha}
 +
 \kappa_{ab} 
 (\overline{N_{L}'^{c}})_{a} (N_{L}')_{b} \phi 
 +
 \mu
 \varphi^{*}
 H_{d} {\rm i} \tau^{2} H_{u}
 +
 \omega
 \varphi \varphi \phi^{*}
 \nonumber 
 \\
 &\hspace{1.5cm}
 +
 (\overline{N_{R}})_{a}
 M_{ab}
 (N_{L}')_{a}
 +
 {\rm H.c.}
 \Bigr]
 +
 M_{\phi}^{2}
 \phi^{*} \phi
 +
 M_{\varphi}^{2}
 \varphi^{*} \varphi
 +
 \cdots.
\label{equ:L-decon-3-9}
\end{align}
It can be implemented by the following charge assignments under a $\mathbb{Z}_7$ (\cf, \Sec~\ref{sec:eff}):
\begin{equation}
q_{H_{u}} = 0 \, , \quad q_{H_{d}} = 6 \, , \quad q_{L} = 1, \, \quad
q_{N_{R}} = q_{N_{L}'} = 1 \, , 
\quad 
q_{\varphi} = 6 \, , 
\quad
q_{\phi} = 5 \, . 
\end{equation}
If the scalars are integrated out, we obtain the inverse see-saw mass matrix with a $\mu$-term
\begin{align}
\mathscr{L}=
\frac{1}{2}
\begin{pmatrix}
\overline{\nu_{L}^{c}}_{\alpha}
&
\overline{N_{R}}_{a}
&
(\overline{N_{L}'^{c}})_{c}
\end{pmatrix}
\begin{pmatrix}
 0 
 & (Y_{\nu}^{\sf T})_{\alpha b} \langle H_{u}^{0} \rangle 
 & 0 
 \\
 (Y_{\nu})_{a \beta} \langle H_{u}^{0} \rangle 
 & 0 
 & M_{ad} 
 \\
 0 
 & (M^{\sf T})_{cb}
 & (\Lambda^{-3})_{cd} \langle H_{d}^{0} H_{u}^{0} \rangle^{2}
\end{pmatrix}
\begin{pmatrix}
\nu_{L\beta} \\ (N_{R}^{c})_{b} \\ N_{Ld}'
\end{pmatrix}
+{\rm H.c.}
\label{eq:massmatrix-inverseseesaw-dim9}
\end{align}
with
\begin{align}
(\Lambda^{-3})_{ab}
=
 2 \kappa_{ab} 
 \frac{\mu^{2} \omega}{M_{\phi}^{2} M_{\varphi}^{4}} \sim \mathcal{O} \left( \frac{1}{\Lambda_{\mathrm{NP}}^3} \right) \, .
\end{align}
Now the $\mu$-term is suppressed by $\Lambda_{\mathrm{NP}}^{-3}$ and the LNV parameter $\kappa$, \ie, 
extremely small. Neutrino mass, of course, acquires additional suppression from $M$.

\subsection{Two loop generated $\boldsymbol{d=7}$ operator}
\label{sec:loop}

Here we show an example for neutrino mass generation from $\delta
\mathscr{L}_{d=7}^{(2)}$.\footnote{Another realization of the loop
suppressed inverse see-saw with an extended Higgs sector 
is shown in \Ref~\cite{Ma:2009gu}.} This possibility is a very neat example, because all the different concepts from the introduction to reduce the new physics scale are implemented at once: radiative generation of neutrino mass, small LNV parameter, and neutrino mass generation from a higher dimensional operator. 
In all the previous examples, the resulting Lagrangian had a full (new) U(1) symmetry instead of a $\mathbb{Z}_n$, and the U(1) had to be broken by a sector which is independent of neutrino mass. In this example, we will demonstrate that the neutrino mass emerges from the breaking of the U(1) to the $\mathbb{Z}_n$.

We introduce two SM singlet chiral fermions $N_{R}$ and 
$N'_{L}$, and two SM singlet scalars $\phi$ and $S$:
\begin{align} 
\mathscr{L} = & \mathscr{L}_{\rm SM} 
 +
 \Bigl[
 (Y_{N})_{a \alpha}
 (\overline{N_{R}})_{a} H_{d}^{\dagger} L_{\alpha}
 +
 (\alpha_{1})_{a b} \phi 
 (\overline{N_{R}^{c}})_{a} (N_{R})_{b}
 +
 (\alpha_{2})_{ab} \phi 
 (\overline{N_{L}'^{c}})_{a} (N_{L}')_{a}
 \nonumber
 \\
 &
 \hspace{1.5cm}
 +
 \mu S^{*} H_{d} {\rm i} \tau^{2} H_{u}
 +
 (\overline{N_{R}})_{a} M_{ab} (N_{L}')_{b}
 +
 {\rm H.c.}
 \Bigr]
 - 
 \mathscr{V}(H_u,\, H_d,\, \phi,\, S) \, .\label{equ:model-loop}
\end{align}
The relevant part of the scalar potential is given by
\be \label{equ:pot}
\mathscr{V}(H_u,\, H_d,\, \phi,\, S) = 
\left[
\lambda_{1} S \phi^{3}
+
\mu_{1} S^{*} \phi^{2}
+
\lambda_{2} S^{3} \phi^{*}
+
{\rm H.c.}
\right]
+
M_{S}^{2} S^{*} S
+
M_{\phi}^{2} \phi^{*} \phi
+
\cdots
\,. 
\ee
Let us focus on the terms in the bracket in \equ{model-loop}: These
terms respect three independent U(1) symmetries, which can be identified
with hypercharge, lepton number and an additional (new) U(1)
symmetry.
Since lepton number is conserved in this sector of the Lagrangian, no operator can be written for neutrino masses. On the other hand, the scalar potential in \equ{pot} violates all continuous symmetries but hypercharge, while respecting $\mathbb{Z}_5$.  Neutrino masses are therefore only allowed in the presence of the scalar potential, which violates lepton number and the new U(1).
In fact, the scalar potential in \equ{pot} just generates the effective U(1) breaking term in \equ{phi5} after integrating out the $S$ field.

In the following, we choose the $\mathbb{Z}_5$ charges
\begin{equation}
q_{H_{u}} = 0 \, , \quad q_{H_{d}} = 1 \, , \quad q_{L} = 2, \, \quad
q_{N_{R}} = q_{N_L'} = 1 \, , 
\quad
q_{\phi} = 3 \, , 
\quad
q_{S} = 1 \, 
\end{equation}
to implement the Lagrangian in \equ{model-loop} leading to neutrino mass from operator \#4 in \Tab~\ref{tab:operators},
while operators \#1 to \#3 are forbidden.

Now the mass matrix for three types of neutral fermions 
can be written as
\begin{align}
\mathscr{L}=
\frac{1}{2}
\begin{pmatrix}
\overline{\nu_{L}^{c}}_{\alpha}
&
\overline{N_{R}}_{a}
&
(\overline{N_{L}'^{c}})_{c}
\end{pmatrix}
\begin{pmatrix}
 m_{\nu \alpha \beta}^{\text{(2-loop)}}
 & {(Y_{N}^{\sf T})_{\alpha b}} \langle H_{d}^{0*} \rangle
 & (\epsilon Y_{\nu}')^{(\text{1-loop}) {\sf T}}_{\alpha d}
 \\
 {(Y_{N})_{a\beta}}\langle H_{d}^{0*} \rangle
 & \mu'^{(\text{tree})}_{ab}
 & M_{ad}
 \\
 (\epsilon Y_{\nu}')^{\text{(1-loop)}}_{c \beta}
 & {(M^{\sf T})}_{cb} 
 & \mu^{\text{(tree)}}_{cd}
\end{pmatrix}
\begin{pmatrix}
\nu_{L\beta} \\ (N_{R}^{c})_{b} \\ N'_{Ld}
\end{pmatrix}
+{\rm H.c.}.
\end{align}
Assuming that
\be
M_S \sim M_\phi \sim \mu \sim M \equiv \Lambda_{\rm NP} \, 
\ee
we can estimate the elements as
\begin{align}
m_{\nu\alpha \beta}^{\text{(2-loop)}}
\sim&
\frac{1}{(16\pi^{2})^{2}}
\frac{v_{d} v_{u}^{3}}{\Lambda_{\rm
 NP}}
\lambda_{2}
[Y_{N}^{\sf T} (M^{\sf T})^{-1} \alpha_{2} M^{-1} Y_{N}]_{\alpha \beta},
\\
(\epsilon Y_{\nu}')^{\text{(1-loop)}}_{c \beta}
\sim&
\frac{1}{16\pi^{2}}
\frac{v_{d}^{2} v_{u}^{3}}{\Lambda_{\rm NP}^{3}}
\lambda_{2}
[\alpha_{2} M^{-1} Y_{N}]_{c \beta},
\\
\mu^{\text{(tree)}}_{cd}
\sim&
\frac{v_{d}^{3} v_{u}^{3}}{\Lambda_{\rm NP}^{5}} \lambda_{2} 
 (\alpha_{2})_{cd}, \label{m-Maj}
\\
\mu'^{\text{(tree)}}_{ab}
\sim&
\frac{v_{d}^{3} v_{u}^{3}}{\Lambda_{\rm NP}^{5}} \lambda_{2}^{*} 
 (\alpha_{1}^{*})_{ab}. \label{m-Maj'}
\end{align}
The two-loop contribution to neutrino masses comes from the diagram
shown in \figu{dec-dim9} (d).
The one-loop contribution to $\epsilon$-term can be obtain 
by cutting a propagator of $H_{d}$
in \figu{dec-dim9} (d) and giving the VEVs to the 
end of the cut propagator.
Assuming the parameters $\alpha_2$ and $\lambda_2$ are $\mathcal{O}(1)$,
we find that in order to obtain neutrino masses of the order 
of the eV if $\Lambda_{\rm NP} \sim 10 \, \mathrm{TeV}$. 

Notice that the order of magnitude result of Eqs. \eqref{m-Maj} and \eqref{m-Maj'} takes into account the fact that the scalars  $S$ and $\phi$ acquire VEVs, after electroweak symmetry breaking, due to the mixed terms in the Lagrangian.  These VEVs can be estimated at tree level by minimizing the potential. In particular, the VEV of the $\phi$, which contributes to the Majorana masses of the heavy neutrinos, is consistent with the formulas above.

Two things are different in this scenario from those previously considered and are worth stressing. First of all, postulating that neutrino mass is generated from the breaking of the (new) U(1) symmetry down to $\mathbb{Z}_5$ has forced us to consider loop diagrams in order to generate neutrino masses. These introduce an extra suppression by about two orders of magnitude. Secondly, the neutrino masses must come proportional to the couplings in the scalar sector $\alpha_1$ and $\lambda_2$ since they are responsible for LNV and the U(1) breaking. If these couplings are perturbative, they can easily account for some more suppression while still being natural. That is, this model predicts neutrino masses at a scale of new physics that is naturally the TeV scale with large Yukawa couplings. Since flavor violating processes  appear at tree level as $d=6$ operators, we expect new physics within the reach of near future experiments.

A mored detailed discussion is beyond the scope of the current study.

\section{Summary and conclusions}

In this study, it has been demonstrated that neutrino mass can be generated from higher than $d=5$ operators even at tree level if the Lagrangian is charged under a new U(1) or discrete symmetry  to forbid the lower dimensional operators. It has been shown that new scalar fields need to be added in order to realize such mechanisms. We have focused on a Two Higgs Doublet Model (THDM) extension of the Standard Model (SM). The use of higher dimensional operators has  allowed us  to lower the new physics scale to close to the TeV scale, while having ``natural'' Yukawa couplings.

Tree level decompositions of the $d=7$ operator in the THDM have been
discussed, which have led to implementations of the inverse see-saw
mechanism. In this case, the smallness of the lepton number violating
mass term emerges naturally as $\mathcal{O}(v^{2}/\Lambda_{\mathrm{NP}})$. Furthermore, both the $(3,3)$-element and the $(1,3)$-element of the inverse see-saw mass matrix can be easily generated independently, depending on the mediators used.

All possible tree level decompositions of the $d=7$ effective operator $LLH_{u} H_{u} H_{d} H_{u}$ have been systematically studied to find possibilities for natural generalizations of the type~I,~II, and~III see-saw scenarios. For example, our inverse see-saw scenarios have turned out to be two of the simplest generalizations of the type~I see-saw mechanism. Even simpler generalizations with only two extra fields have been found for the type~II see-saw.

It has been also demonstrated that an even stronger suppression of neutrino mass can be achieved by $d>7$ effective operators generated by tree-level-exchanges, or loop-generated $d=7$ operators. One of our examples combines all three elements of modern TeV-scale see-saws: radiative generation of neutrino mass, small lepton number violation, and neutrino mass generation from  a dimension seven operator. In this case, neutrino mass is proportional to the breaking of a continuous new U(1) symmetry to a discrete $\mathbb{Z}_5$. Even with order one couplings, the new physics scale is at a few TeV in this example.

We conclude that there may be plenty of possibilities to implement a TeV scale see-saw mechanism generating higher than $d=5$ effective operators as leading contribution to neutrino mass, which we have studied within the framework of the THDM. The most promising of the UV completions of the effective operators lead to inverse see-saw scenarios with a natural explanation for the smallness of the LNV parameter. Our mechanism can also be applied to SUSY models, which will be studied elsewhere~\cite{Gavela:prep}.

\subsubsection*{Acknowledgments}

We are indebted to Belen Gavela to many interesting physics discussions and for numerous
useful comments.
DH received partial support from CICYT through the project FPA2006-05423, as well as from the Comunidad Aut\'onoma de Madrid through Proyecto HEPHACOS; P-ESP-00346. DH also acknowledges financial support from the MEC through FPU grant AP20053603. 
FB thanks the Departamento de F\'isica Te\'orica de Madrid
(UAM/CSIC) for its hospitality during the completion of this work.
TO and WW would like to acknowledge support from the 
Emmy Noether program of Deutsche Forschungsgemeinschaft, 
contract WI 2639/2-1.

\end{document}